\pdfoutput=1

\documentclass[11pt,twoside,a4paper,cmspaper,final,collab]{cms-tdr}
\def\svnVersion{389851}\def\cmsCernNoTag{CERN-EP/2016-103}\def\cmsCernDate{\today}\def\cmsMessage{Published in Physics Letters B as \href{http://dx.doi.org/10.1016/j.physletb.2017.02.007}{\doi{10.1016/j.physletb.2017.02.007.}}}

\begin{document}\cmsNoteHeader{SUS-14-006}

\hyphenation{had-ron-i-za-tion}
\hyphenation{cal-or-i-me-ter}
\hyphenation{de-vices}
\RCS$HeadURL: svn+ssh://svn.cern.ch/reps/tdr2/papers/SUS-14-006/trunk/SUS-14-006.tex $
\RCS$Id: SUS-14-006.tex 387983 2017-02-17 21:19:22Z bainbrid $

\newlength\cmsFigWidth
\ifthenelse{\boolean{cms@external}}{\setlength\cmsFigWidth{0.98\columnwidth}}{\setlength\cmsFigWidth{0.8\textwidth}}
\ifthenelse{\boolean{cms@external}}{\providecommand{\cmsLeft}{top}}{\providecommand{\cmsLeft}{left}}
\ifthenelse{\boolean{cms@external}}{\providecommand{\cmsRight}{bottom}}{\providecommand{\cmsRight}{right}}

\newcommand{\cls}{\text{CL$_s$}\xspace}
\newcommand{\scalht}{\ensuremath{H_\text{T}}\xspace}
\newcommand{\HTmiss}{\ensuremath{H_\text{T}^\text{miss}}\xspace}
\newcommand{\dht}{\ensuremath{\Delta\scalht}\xspace}
\newcommand{\alphat}{\ensuremath{\alpha_\mathrm{T}}\xspace}
\newcommand{\njet}{\ensuremath{N_{\text{jet}}}\xspace}
\newcommand{\njetlow}{\ensuremath{2 \leq \njet \leq 3}\xspace}
\newcommand{\njethigh}{\ensuremath{\njet \geq 4}\xspace}
\newcommand{\nb}{\ensuremath{N_{\PQb}}\xspace}
\newcommand{\mj}{\ensuremath{\mu + \text{jets}}\xspace}
\newcommand{\mmj}{\ensuremath{\mu\mu + \text{jets}}\xspace}
\newcommand{\gj}{\ensuremath{\gamma + \text{jets}}\xspace}
\newcommand{\wjets}{\ensuremath{\PW + \text{jets}}\xspace}
\newcommand{\wmujets}{\ensuremath{\PW \to \mu\nu + \text{jets}}\xspace}
\newcommand{\zjets}{\ensuremath{\cPZ + \text{jets}}\xspace}
\newcommand{\znunujets}{\ensuremath{\cPZ \to \cPgn\cPagn + \text{jets}}\xspace}
\newcommand{\znunu}{\ensuremath{\cPZ \to \cPgn\cPagn}\xspace}
\newcommand{\zmumujets}{\ensuremath{\cPZ \to \mu\mu + \text{jets}}\xspace}
\newcommand{\dphi}{\ensuremath{\Delta\phi^{*}_\text{min}}\xspace}
\newcommand{\dm}{\ensuremath{\Delta m}\xspace}
\newcommand{\alphatmin}{\ensuremath{\alphat^\text{min}}\xspace}
\newcommand{\mhtmet}{\ensuremath{\HTmiss / \ETmiss}\xspace}
\newcommand{\ffbp}{\ensuremath{f{\bar{f}'}}\xspace}
\cmsNoteHeader{SUS-14-006}

\title{Search for top squark pair production in
  compressed-mass-spectrum scenarios in proton-proton collisions at
  $\sqrt{s} = 8\TeV$ using the \alphat variable}

\date{\today}

\abstract{An inclusive search is performed for supersymmetry in final
  states containing jets and an apparent imbalance in transverse
  momentum, \ptvecmiss, due to the production of unobserved weakly
  interacting particles in pp collisions at a centre-of-mass energy of
  8\TeV. The data, recorded with the CMS detector at the CERN LHC,
  correspond to an integrated luminosity of 18.5\fbinv. The
  dimensionless kinematic variable \alphat is used to discriminate
  between events with genuine \ptvecmiss associated with unobserved
  particles and spurious values of \ptvecmiss arising from jet energy
  mismeasurements. No excess of event yields above the expected
  standard model backgrounds is observed. The results are interpreted
  in terms of constraints on the parameter space of several simplified
  models of supersymmetry that assume the pair production of top
  squarks. The search provides sensitivity to a broad range of top
  squark ($\PSQt$) decay modes, including the two-body decay $\PSQt
  \to \PQc \PSGczDo$, where \PQc is a charm quark and $\PSGczDo$ is
  the lightest neutralino, as well as the four-body decay $\PSQt \to
  \PQb \ffbp \PSGczDo$, where \PQb is a bottom quark and $f$ and
  $\bar{f}'$ are fermions produced in the decay of an intermediate
  off-shell W boson. These modes dominate in scenarios in which the
  top squark and lightest neutralino are nearly degenerate in
  mass. For these modes, top squarks with masses as large as 260 and
  225\GeV are excluded, respectively, for the two- and four-body
  decays.}

\hypersetup{
  pdfauthor={CMS Collaboration},
  pdftitle={Search for top squark pair production in
    compressed-mass-spectrum scenarios in proton-proton collisions at
    sqrt(s) = 8 TeV using the alphaT variable},
  pdfsubject={CMS},
  pdfkeywords={CMS, physics, SUSY, jets, missing transverse momentum,
    alphaT}
}

\maketitle

\section{Introduction}

The standard model (SM) is widely regarded as an effective
approximation, valid at low energies, of a more complete theory of
particle interactions, such as supersymmetry (SUSY)~\cite{ref:SUSY-1,
  ref:SUSY0, ref:SUSY1, ref:SUSY2, ref:SUSY3, ref:SUSY4,
  ref:hierarchy1, ref:hierarchy2}, which would supersede the SM at
higher energy scales. A realisation of SUSY with TeV-scale
third-generation squarks is motivated by the cancellation of
quadratically divergent loop corrections to the mass of the Higgs
boson~\cite{ref:atlashiggsdiscovery, ref:cmshiggsdiscoverylong}
avoiding the need for significant fine tuning~\cite{ref:hierarchy1,
  ref:hierarchy2, ref:barbierinsusy}. In R-parity-conserving
SUSY~\cite{Farrar:1978xj}, supersymmetric particles (sparticles) such
as squarks and gluinos are produced in pairs and decay to the lightest
stable supersymmetric particle (LSP), which is generally assumed to be
a weakly interacting and massive neutralino, $\PSGczDo$. A
characteristic signature of these events is a final state with jets
accompanied by an apparent, significant imbalance in transverse
momentum, \ptvecmiss, due to unobserved $\PSGczDo$ particles that can
carry substantial momentum.

The lack of evidence to date for SUSY at the CERN LHC has led to the
careful consideration of regions of the SUSY parameter space that have
a relatively weak coverage in the experimental programme.  One such
class of models is that of compressed mass spectra, in which the LSP
lies close in mass to the parent sparticle produced in the collisions.
Models in which both the top squark ($\PSQt$) and neutralino LSP are
light and nearly degenerate in mass are phenomenologically well
motivated~\cite{Boehm:1999tr,Boehm:1999bj,Balazs:2004bu,
  Martin:2007gf,
  Martin:2007hn,Carena:2008mj,Grober:2014aha,Grober:2015fia}. For a
mass splitting $\dm = m_{\PSQt} - m_{\PSGczDo} < m_{\PW}$, where
$m_{\PW}$ is the mass of the W boson, the decay modes available to the
top squark are either loop-induced, flavour-changing neutral current
decays to a charm (c) quark and a neutralino, $\PSQt \to
\PQc\PSGczDo$, or four-body decays, $\PSQt \to {\PQb \ffbp} \PSGczDo$,
where b is a bottom quark with $f$ and $\bar{f}'$ fermions from, for
example, an off-shell W boson decay. Improved experimental acceptance
for systems with compressed mass spectra can be achieved by requiring
the sparticles to be produced in association with jets from
initial-state radiation (ISR). The sparticle decay products from these
systems can be Lorentz boosted to values of transverse momentum \pt
within the experimental acceptance if they recoil against a
sufficiently high-\pt jet from ISR. This topology is exploited by
searches that consider $\text{``monojet''} + \ptvecmiss$ final
states~\cite{atlas-13, atlas-6, cms-9}. The reliance on ISR is reduced
for systems with larger \dm, as in this case the sparticle decay
products can have sufficiently large values of \pt to lie within the
experimental acceptance even without the Lorentz boost from ISR.

This letter presents an inclusive search for the pair production of
massive coloured sparticles in final states with two or more energetic
jets and \ptvecmiss in pp collisions at $\sqrt{s} = 8\TeV$. The data
correspond to an integrated luminosity of $18.5 \pm
0.5\fbinv$~\cite{lumi} collected with the CMS detector at the LHC. The
search is based upon a kinematic variable \alphat, described in
Section~\ref{sec:alphat}, which offers powerful discrimination against
SM multijet production, and adheres to a strategy of maximising
experimental acceptance through the application of loose selection
requirements to provide sensitivity to a wide range of SUSY
models. Previous versions of this search were reported at $\sqrt{s} =
7\TeV$~\cite{RA1Paper, RA1Paper2011, RA1Paper2011FULL}, and for an
initial sample of data corresponding to 11.7\fbinv at
$8\TeV$~\cite{RA1Paper2012}. Other LHC searches for manifestations of
SUSY in all-jet final states are presented in Refs.~\cite{atlas-0,
  atlas-1, atlas-2, atlas-3, atlas-4, atlas-5, atlas-11, atlas-7,
  atlas-8, atlas-9, atlas-10, atlas-12, Aad:2016eki, Aaboud:2016zdn,
  cms-1, cms-2, cms-3, cms-4, cms-8,cms-11,cms-5, cms-6, cms-7,
  cms-10, cms-12, cms-13, atlas-13, atlas-6, cms-9}. Recent searches
for top squark production in leptonic final states can be found in
Refs.~\cite{Aad:2015pfx} (and references therein)
and~\cite{Khachatryan:2016pup, Aaboud:2016lwz}.

The search makes use of the number of reconstructed jets per event
(\njet), the number of these jets identified as originating from b
quarks (\nb), and the sum of the transverse energies of these jets
(\scalht), where the transverse energy of a jet is given by $\ET =
E\sin\theta$, with $E$ the energy of the jet and $\theta$ its polar
angle with respect to the beam axis. The three discriminants provide
sensitivity to different production mechanisms of massive coloured
sparticles at hadron colliders (\ie squark-squark, squark-gluino, and
gluino-gluino), to a large range of mass splittings between the parent
sparticle and the LSP, and to third-generation squark signatures.
While the search results can be interpreted with a broad range of
models involving the strong production of coloured sparticles leading
to final states with both low and high b quark content, we focus on
the parameter space of simplified models~\cite{Alwall:2008ag,
  Alwall:2008va, sms} that assumes the pair production of top squarks,
including the nearly mass-degenerate scenarios described
above. Furthermore, interpretations are provided for top squarks that
decay to the $\PSGczDo$ either directly in association with a top
quark ($\PSQt \to \PQt \PSGczDo$), or via an intermediate lightest
chargino $\PSGcpmDo$ in association with a bottom quark, with the
subsequent decay of the $\PSGcpmDo$ to the $\PSGczDo$ and a W boson
($\PSQt \to \PQb \PSGcpmDo \to {\PQb\PW}^{\pm(*)} \PSGczDo$). All
models assume only the pair production of the low-mass eigenstate
$\PSQt_1$, with the $\PSQt_2$ decoupled to a high mass.

Several aspects of the present search are improved relative to the
results of Ref.~\cite{RA1Paper2012} in order to increase the
sensitivity to models with nearly mass-degenerate $\PSQt$ and
$\PSGczDo$ states. The signal region is extended to incorporate events
with a low level of jet activity using a parked data set collected
with a dedicated trigger stream~\cite{CMS:2012ooa}, where ``parked''
means that, due to limitations in the available processing capability,
the data were recorded without being processed through the
reconstruction software, and were processed only subsequent to the end
of the 2012 data collection period. Furthermore, tight requirements on
a combination of kinematic variables are employed to suppress multijet
production to the sub-percent level relative to the total remaining
number of background events from other SM processes. Finally, an event
veto based on isolated tracks is used to further suppress SM
background contributions from $\tau \to \text{hadrons} + \nu$ decays
and misreconstructed electrons and muons. These features yield an
increased experimental acceptance to events with low jet activity, and
improvements in the control of SM backgrounds, which are crucial for
enhancing sensitivity to new sources of physics with nearly degenerate
mass spectra.

\section{The CMS detector}

The central feature of the CMS detector is a superconducting solenoid
providing an axial magnetic field of 3.8\unit{T}. The CMS detector is
nearly hermetic, which allows for accurate momentum balance
measurements in the plane transverse to the beam axis.

Charged particle trajectories are measured by a silicon pixel and
strip tracker system, with full azimuthal ($\phi$) coverage and a
pseudo-rapidity acceptance $\abs{\eta} < 2.5$.  Isolated particles of
\pt = 100\GeV emitted at $\abs{\eta} < 1.4$ have track resolutions of
2.8\% in \pt and 10 (30)\mum in the transverse (longitudinal) impact
parameter \cite{TRK-11-001}.

A lead tungstate crystal electromagnetic calorimeter (ECAL) and a
brass and scintillator hadron calorimeter (HCAL) surround the tracking
volume and provide coverage over $\abs{\eta} < 3.0$. A forward HCAL
extends the coverage to $\abs{\eta} < 5.0$. In the barrel section of
the ECAL, an energy resolution of about 1\% is achieved for
unconverted or late-converting photons with energies on the order of
several tens of GeV. In the $\eta$--$\phi$ plane, and for $\abs{\eta}<
1.48$, the HCAL cells map onto $5 \times 5$ arrays of ECAL crystals to
form calorimeter towers projecting radially outwards from a location
near the nominal interaction point. At larger values of $\abs{ \eta
}$, the size of the towers increases and the matching ECAL arrays
contain fewer crystals. Within each tower, the energy deposits in ECAL
and HCAL cells are summed to define the calorimeter tower energies,
subsequently used to provide the energies and directions of
reconstructed jets. The HCAL, when combined with the ECAL, measures
jet energies with a resolution of approximately 40\% at 12\GeV, 5\% at
100\GeV, and 4\% at 1\TeV.

Muons are identified in gas ionisation detectors embedded in the steel
flux-return yoke of the magnet. Muons are measured in the range
$\abs{\eta}< 2.4$. By matching track segments reconstructed in the
muon detectors to segments measured in the silicon tracker, a relative
transverse momentum resolution of 1.3--2.0\% and $<$10\% is achieved
for muons with, respectively, $20 <\pt < 100\GeV$ and $\pt <
1\TeV$~\cite{Chatrchyan:2012xi}.

The first level (L1) of the CMS trigger system, composed of custom
hardware processors, uses information from the calorimeters and muon
detectors to select events of interest within a fixed time interval of
less than 4\mus. The high-level trigger (HLT) processor farm further
decreases the event rate from around 100\unit{kHz} to about
600\unit{Hz}, before data storage. Of these events, about half are
reconstructed promptly. The other half represent the parked data set
referred to above.

A more detailed description of the CMS detector, together with a
definition of the coordinate system used and the relevant kinematic
variables, can be found in Ref.~\cite{Chatrchyan:2008zzk}.

\section{The \texorpdfstring{\alphat}{AlphaT} variable\label{sec:alphat}}

The \alphat kinematic variable, first introduced in
Refs.~\cite{Randall:2008rw, RA1Paper}, is used to efficiently reject
events that do not contain significant \ptvecmiss or that contain
large \ptvecmiss only because of transverse momentum mismeasurements,
while retaining sensitivity to new-physics events with significant
\ptvecmiss. The \alphat variable depends solely on the transverse
energies and azimuthal angles of jets, and is intrinsically robust
against the presence of jet energy mismeasurements in multijet
systems.

For events containing only two jets, \alphat is defined as $\alphat =
\ET^{\text j_2}/M_\mathrm{T}$, where $\ET^{\text j_2}$ is the
transverse energy of the jet with smaller \ET, and $M_\mathrm{T}$ is
the transverse mass of the dijet system, defined as:
\begin{equation}
  \label{eq:mt}
  M_\mathrm{T} = \sqrt{ \left( \sum_{i=1}^2 \ET^{\mathrm{j}_i}
    \right)^2 - \left( \sum_{i=1}^2 p_x^{\mathrm{j}_i} \right)^2 - \left(
      \sum_{i=1}^2 p_y^{\mathrm{j}_i} \right)^2},
\end{equation}
where $\ET^{\mathrm{j}_i}$, $p_x^{\mathrm{j}_i}$, and
$p_y^{\mathrm{j}_i}$ are, respectively, the transverse energy and $x$
or $y$ components of the transverse momentum of jet $\mathrm{j}_i$.
For a perfectly measured dijet event with $\ET^{\mathrm{j}_1} =
\ET^{\mathrm{j}_2}$ and the jets in the back-to-back configuration
($\Delta\phi = \pi$), and in the limit in which the momentum of each
jet is large compared with its mass, the value of \alphat is 0.5.  For
an imbalance in the \ET values of the two back-to-back jets, whether
due to an over- or under-measurement of the \ET of either jet, then
$\ET^{\mathrm{j}_2} < 0.5M_\mathrm{T}$. This in turn implies $\alphat
< 0.5$, giving the variable its intrinsic robustness. Values of
\alphat significantly greater than 0.5 are observed when the two jets
are not back-to-back and recoil against significant, genuine
$\ptvecmiss$ from weakly interacting particles that escape the
detector, such as neutrinos.

The definition of the \alphat variable can be generalised for events
with more than two jets~\cite{RA1Paper}. The mass scale for any
process is characterised through the scalar \ET sum of jets, defined
as $\scalht = \sum_{i=1}^{N_\text{jet}} \ET^{\mathrm{j}_i}$, where
$N_\text{jet}$ is the number of jets with \ET above a predefined
threshold. The estimator for $\abs{\ptvecmiss}$ is given by the
magnitude of the vector \pt sum of all the jets, defined by $\HTmiss =
\abs{\sum_{i=1}^{N_\text{jet}} \vec{\pt}^{\mathrm{j}_i}}$. For events
with three or more jets, a pseudo-dijet system is formed by combining
the jets in the event into two pseudo-jets. The total \scalht for each
of the two pseudo-jets is given by the scalar \ET sum of its
contributing jets. The combination chosen is the one that minimises
\dht, defined as the difference between the \scalht of the two
pseudo-jets. This clustering criterion assumes a balanced-momentum
hypothesis, $\abs{\ptvecmiss} \approx 0\GeV$, which provides the best
separation between SM multijet events and events with genuine
\ptvecmiss. The \alphat definition can then be generalised to:

\begin{equation}
  \label{eq:alphat}
  \alphat = \frac{1}{2} \frac{\scalht -
    \dht}{\sqrt{(\scalht)^2 - (\HTmiss)^2}}.
\end{equation}

When jet energies are mismeasured, or there are neutrinos from
heavy-flavour quark decays, the magnitude of \HTmiss and \dht are
highly correlated. This correlation is much weaker for
R-parity-conserving SUSY events, where each of the two decay chains
produces an undetected LSP.

\section{Event reconstruction and selection}
\label{sec:selections}

The event reconstruction and selection criteria described below are
discussed in greater detail in Ref.~\cite{RA1Paper2012}. To suppress
SM processes with genuine \ptvecmiss from neutrinos, events containing
an isolated electron~\cite{Khachatryan:2015hwa} or
muon~\cite{Chatrchyan:2012xi} with $\pt > 10\GeV$ are
vetoed. Furthermore, events containing an isolated
track~\cite{single-lepton-stop} with $\pt > 10\GeV$ are vetoed. Events
containing isolated photons~\cite{Khachatryan:2015iwa} with $\pt >
25\GeV$ are also vetoed to ensure an event sample comprising only
multijet final states.

Jets are reconstructed from the energy deposits in the calorimeter
towers, clustered using the anti-\kt algorithm~\cite{antikt} with a
radius parameter of 0.5. The jet energies measured in the calorimeters
are corrected to account for multiple pp interactions within an event
(pileup), and to establish a uniform relative response in $\eta$ and a
calibrated absolute response in \pt~\cite{Chatrchyan:2011ds}.  Jets
are identified as originating from b quarks using the ``medium''
working point of the combined secondary vertex
algorithm~\cite{Chatrchyan:2012jua}, such that the probability to
misidentify jets originating from light-flavour partons (gluons and u,
d, or s quarks) as b quark jets is approximately 1\% for jets with
$\pt = 80\GeV$. The ``medium'' working point results in a b-tagging
efficiency, \ie the probability to correctly identify jets as
originating from b quarks, in the range 60--70\% depending on the jet
\pt.

All jets are required to satisfy $\abs{\eta} < 3.0$, and the jet with
largest \ET is also required to satisfy $\abs{\eta} < 2.5$. All jets
and the two jets with largest \ET are, respectively, subjected to a
nominal ($\ET > 50\GeV$) and higher ($\ET > 100\GeV$) threshold.
Events are required to contain at least two jets that satisfy the
aforementioned $\ET$ and $\eta$ requirements. The value of \scalht for
each event is determined from these jets. If $\scalht < 375\GeV$, the
respective jet \ET thresholds are lowered to 43 and 87\GeV, \scalht is
recalculated, and the event is reconsidered for selection. If the
recalculated \scalht is less than 325\GeV, the respective \ET
thresholds are lowered yet further, to 37 and 73\GeV and \scalht again
recalculated.  If this newly recalculated \scalht is less than
200\GeV, the event is rejected. The scheme is summarised in
Table~\ref{tab:thresholds}. Events can be selected with this iterative
procedure even if they do not satisfy the sets of tighter requirements
on the \ET thresholds. The reason why lower jet \ET thresholds are
employed for $200 < \scalht < 375\GeV$ is to maintain a similar
background composition in all \scalht bins, and to increase the
acceptance for SUSY models characterised by compressed mass
spectra. Significant jet activity in the event is established by
requiring $\scalht > 200\GeV$, which also ensures high efficiency for
the trigger conditions, described below, used to record the
events. Events are vetoed if rare, anomalous signals are identified in
the calorimeters~\cite{Chatrchyan:2009hy} or if any jet satisfies $\ET
> 50\GeV$ and has $\abs{\eta} > 3$, in order to enhance the
performance of \HTmiss as an estimator of $\abs{\ptvecmiss}$.

\begin{table*}[!htb]
  \topcaption{\scalht-dependent thresholds on the \ET values of jets and
    \alphat values.\label{tab:thresholds}}
  \centering
\newcolumntype{.}{D{.}{.}{2}}
  \begin{tabular}{ l.... }
    \hline
    \scalht (\GeVns)                 & \multicolumn{1}{c}{200--275}      & \multicolumn{1}{c}{275--325}      & \multicolumn{1}{c}{325--375}      & \multicolumn{1}{c}{$>$375}        \\
    \hline
    Highest \ET jet (\GeVns)         & 73            & 73            & 87            & 100           \\
    Next-to-highest \ET jet (\GeVns) & 73            & 73            & 87            & 100           \\
    \ET of other jets (\GeVns)       & 37            & 37            & 43            & 50            \\
    \alphat                       & 0.65          & 0.60          & 0.55          & 0.55          \\
    \hline
  \end{tabular}
\end{table*}

Events are categorised according to the number of jets per event,
\njetlow or \njethigh, and the number of reconstructed b quark jets
per event, $\nb = 0$, 1, 2, 3, or $\geq$4.  For events containing
exactly zero or one b quark jet, we employ eleven bins in \scalht:
three bins at low jet activity in the range of $200 < \scalht <
375\GeV$, as detailed in Table~\ref{tab:thresholds}, an additional
seven bins $100\GeV$ wide in the range of $375 < \scalht < 1075\GeV$,
and an open final bin $\scalht > 1075\GeV$.  For events containing two
or three (at least four) b quark jets, a total of nine (four) bins are
used in \scalht, with an open final bin $\scalht > 875\, (375)\GeV$.
This categorisation according to \njet, \nb, and \scalht results in a
total of eight (\njet,\nb) event categories and 75
bins. An overview of the binning scheme is provided by
Table~\ref{tab:fit-result}.

For events satisfying the above selection criteria, the multijet
background dominates over all other SM sources. Multijet events
populate the region $\alphat \lesssim 0.5$, and the \alphat
distribution is characterised by a sharp edge at 0.5, beyond which the
multijet event yield falls by several orders of magnitude. Multijet
events with extremely rare but large stochastic fluctuations in the
calorimetric measurements of jet energies can lead to values of
\alphat slightly above 0.5. The edge at 0.5 sharpens with increasing
\scalht for multijet events, primarily due to a corresponding increase
in the average jet energy and a consequent improvement in the jet
energy resolution. The contribution from multijet events is suppressed
by more than five orders of magnitude by imposing the
\scalht-dependent \alphat requirements summarised in
Table~\ref{tab:thresholds}.

Several beam- and detector-related effects, such as interactions from
beam halo, reconstruction failures, detector noise, or event
misreconstruction due to detector inefficiencies, can lead to events
with large, unphysical values of \ptvecmiss and values of \alphat 
greater than 0.55. These types of events are rejected with high
efficiency by applying a range of vetoes~\cite{cms-met}.

Two final event vetoes complete the definition of the signal region.
An estimator for \ptvecmiss is defined by the negative of the vector
sum of the transverse momenta of all reconstructed particles in an
event, as determined by the particle-flow (PF)
algorithm~\cite{CMS-PAS-PFT-09-001, CMS-PAS-PFT-10-001}. The magnitude
of this vectorial summation is referred to as \ETmiss. The first veto
concerns the rare circumstance in which several jets, collinear in
$\phi$ and each with \pt below its respective threshold, result in
significant \HTmiss. This type of background, typical of multijet
events, is suppressed while maintaining high efficiency for SM or
new-physics processes with genuine \ptvecmiss by requiring $\HTmiss /
\MET < 1.25$. The second veto considers the minimum azimuthal
separation between a jet and the negative of the vector sum derived
from the transverse momenta of all other jets in the event, which is
referred to as \dphi~\cite{RA1Paper}. This variable is employed to
suppress potential contributions from energetic multijet events that
have significant \ptvecmiss through the production of neutrinos in
semileptonic heavy-flavour decays. Such neutrinos are typically
collinear with the axis of a jet. We impose the requirement $\dphi >
0.3$, which effectively suppresses this background as determined using
control data.

\section{Triggers and data control samples \label{sec:triggers}}

Candidate signal events are recorded under multiple jet-based trigger
conditions that require both \scalht and \alphat to satisfy
predetermined thresholds. The trigger-level jet energies are corrected
to account for energy scale and pileup effects. The trigger
efficiencies for the SM backgrounds are measured using a sample of \mj
events, which provides an unbiased coverage of the kinematic phase
space when the muon is ignored. The efficiencies are determined as a
function of \njet and \scalht, and lie in the range 79--98\% and
$>$99\% for $200 < \scalht < 375\GeV$ and $\scalht > 375\GeV$,
respectively. The inefficiencies at low values of \scalht, which are
accounted for in the final result, arise from conditions imposed on L1
trigger quantities. Statistical uncertainties of a few percent are
considered. Simulation-based studies demonstrate that trigger
inefficiencies for signal events are typically negligible.

A set of prescaled $\scalht$ trigger conditions is used to record
events for a multijet-enriched control sample, defined by relaxed
requirements on \alphat, \dphi, and \mhtmet with respect to the signal
region. This event sample is used to estimate the multijet background
contribution.

Significant background in the signal region is expected from SM
processes with genuine \ptvecmiss in the final state. The dominant
processes are the associated production of W or Z bosons and jets,
with the decays \znunu or $\PW^\pm \to \ell\nu$ ($\ell=\Pe$, $\Pgm$,
$\Pgt$), and top quark pair production followed by semileptonic top
quark decay. Three separate data control regions are used to estimate
the background from these processes. The control regions are defined
through the selection of \mj, \mmj, or \gj
events~\cite{RA1Paper2012}. The selection criteria are chosen such
that the SM processes and their kinematic properties resemble as
closely as possible the SM background behaviour in the signal region,
once the muon, dimuon system, or photon are ignored in the
determination of quantities such as \scalht and \alphat. The event
selection criteria are defined to ensure that the potential
contribution from multijet events or from a wide variety of SUSY
models (\ie so-called signal contamination) is negligible. Events are
categorised according to \njet, \nb, and \scalht, identically to the
scheme used for events in the signal region, as defined in
Section~\ref{sec:selections}.

The \mj sample is recorded using a trigger that requires an isolated
muon. The event selection criteria are chosen so that the trigger is
maximally efficient ($\approx$90\%). Furthermore, the muon is required
to be well separated from the jets in the event, and the transverse
mass formed by the muon and \ETmiss system must lie between 30 and
125\GeV to ensure a sample rich in W bosons (produced promptly or from
the decay of top quarks). The \mmj sample uses the same trigger
condition (efficiency $\approx$99\%) and similar selection criteria as
the \mj sample, specifically requiring two oppositely charged isolated
muons that are well separated from the jets in the event, and with a
dilepton invariant mass within a $\pm 25\GeV$ window around the
nominal mass of the Z boson. For both the muon and dimuon samples, no
requirement is made on \alphat, in order to increase the statistical
precision of the predictions from these samples.  The \gj events are
recorded using a single-photon trigger condition. The event selection
criteria require an isolated photon with $\pt > 165\GeV$, $\scalht >
375\GeV$, and $\alphat > 0.55$, yielding a trigger efficiency
of~$\gtrsim$99\%.

\section{Multijet background suppression \label{sec:multijet}}

The signal region is defined in a manner to suppress the expected
contribution from multijet events to the sub-percent level relative to
the expected background from other SM processes for all event
categories and \scalht bins. This is achieved through very restrictive
requirements on the \alphat and \dphi variables, as described
above. In this section, we discuss these requirements further,
together with the procedure for estimating the remaining multijet
background.

Independent estimates are determined per bin in the signal region,
defined in terms of \njet, \nb, and \scalht. The method utilises the
multijet-enriched control sample introduced in
Section~\ref{sec:triggers}, defined by $0.505 < \alphat < 0.55$ and no
threshold requirements on \dphi or \mhtmet. The event counts in this
data sideband are corrected to account for contamination from
nonmultijet processes, which are estimated using the method described
in Section~\ref{sec:ewk}. The method exploits the evolution of the
ratio $\mathcal{R}(\alphat)$, defined by the number of (corrected)
event counts that satisfy the requirement $\mhtmet < 1.25$ to the
number that fail, as a function of \alphat. The ratio
$\mathcal{R}(\alphat)$ is observed to monotonically fall as a function
of \alphat and is modelled, independently for each bin, with an
exponential function $\mathcal{F}(\alphat)$. An additional
multijet-enriched data sideband, defined by $\mhtmet > 1.25$ and
$\alphat > 0.55$, is used to determine the number of (corrected)
events $\mathcal{N}(\alphat > \alphatmin)$ per bin that satisfy a
minimum threshold requirement on \alphat. Finally, an estimate of the
multijet background for each bin is determined as a function of the
threshold $\alphatmin$ based on the product of $\mathcal{N}(\alphat >
\alphatmin)$ and the extrapolated value of the ratio from the
corresponding fit, $\mathcal{F}(\alphat > \alphatmin)$.

The $\alphat$ value required to suppress the predicted multijet
contribution to the sub-percent level relative to the total SM
background is determined independently for each bin of the signal
region. The $\alphatmin$ thresholds determined from this method are
summarised in Table~\ref{tab:thresholds} and, for simplicity, are
chosen to be identical for all \njet and \nb categories. Higher
\alphat thresholds are required than those used for
Ref.~\cite{RA1Paper2012} because of higher pileup conditions in the
latter half of the data collected in 2012 and because of the addition
of the low \scalht bins.

Various checks are performed in simulation and in data to assure
closure, which, in simulation refers to the ability of the method to
correctly predict the background rates found in simulated data, and,
in data, refers to the consistency between the data-derived
predictions for, and counts in, a separate multijet-enriched
validation sample in data. The exponential functions are found to
adequately model the observed behaviour in data and
simulation. Systematic uncertainties in the predictions are obtained
from the differences observed using alternative fit functions and can
be as large as $\sim$100\%.

Following application of the \alphat requirements, residual
contributions from multijet events with significant \ptvecmiss due to
semileptonic heavy-flavour decays are suppressed by requiring $\dphi >
0.3$, as discussed in Section~\ref{sec:selections}. This suppression
is validated in simulation and in data using a control sample defined
by the requirements $\scalht > 775 \GeV$ and either $0.51 < \alphat <
0.55$ or $\mhtmet > 1.25$.  These events are selected with an
unprescaled \scalht trigger, allowing a study of the performance of
the selection requirements in the low \alphat region around 0.51,
which corresponds to similar \HTmiss values as employed in the lowest
\scalht bins.  From these studies, the remaining multijet background
is found to be at the sub-percent level. With this level of
suppression, any residual contribution from multijet events is assumed
to be negligible compared to the uncertainties associated with the
nonmultijet backgrounds (described below) and is ignored.

\section{Estimation of nonmultijet backgrounds\label{sec:ewk}}

In events with few jets or few b quark jets, the largest backgrounds
are $\znunu\ + \text{jets}$ or $\PW^\pm \to \ell\nu\ +
\text{jets}$. At higher jet or b quark jet multiplicities, \ttbar and
single top production also become an important source of
background. For W boson decays that yield an electron or muon
(possibly originating from leptonic $\Pgt$ decays), the background
arises when the $\Pe$ or $\Pgm$ is not rejected through the dedicated
lepton vetoes. Background also arises when the $\tau$ lepton decays to
neutrinos and hadrons, which are identified as a jet. The veto of
events containing at least one isolated track is efficient at further
suppressing these backgrounds, including those from single-prong
$\tau$-lepton decays, by as much as $\sim$50\% for categories enriched
in \ttbar.

\begin{figure}[t!]
  \centering
  \includegraphics[width=0.49\textwidth]{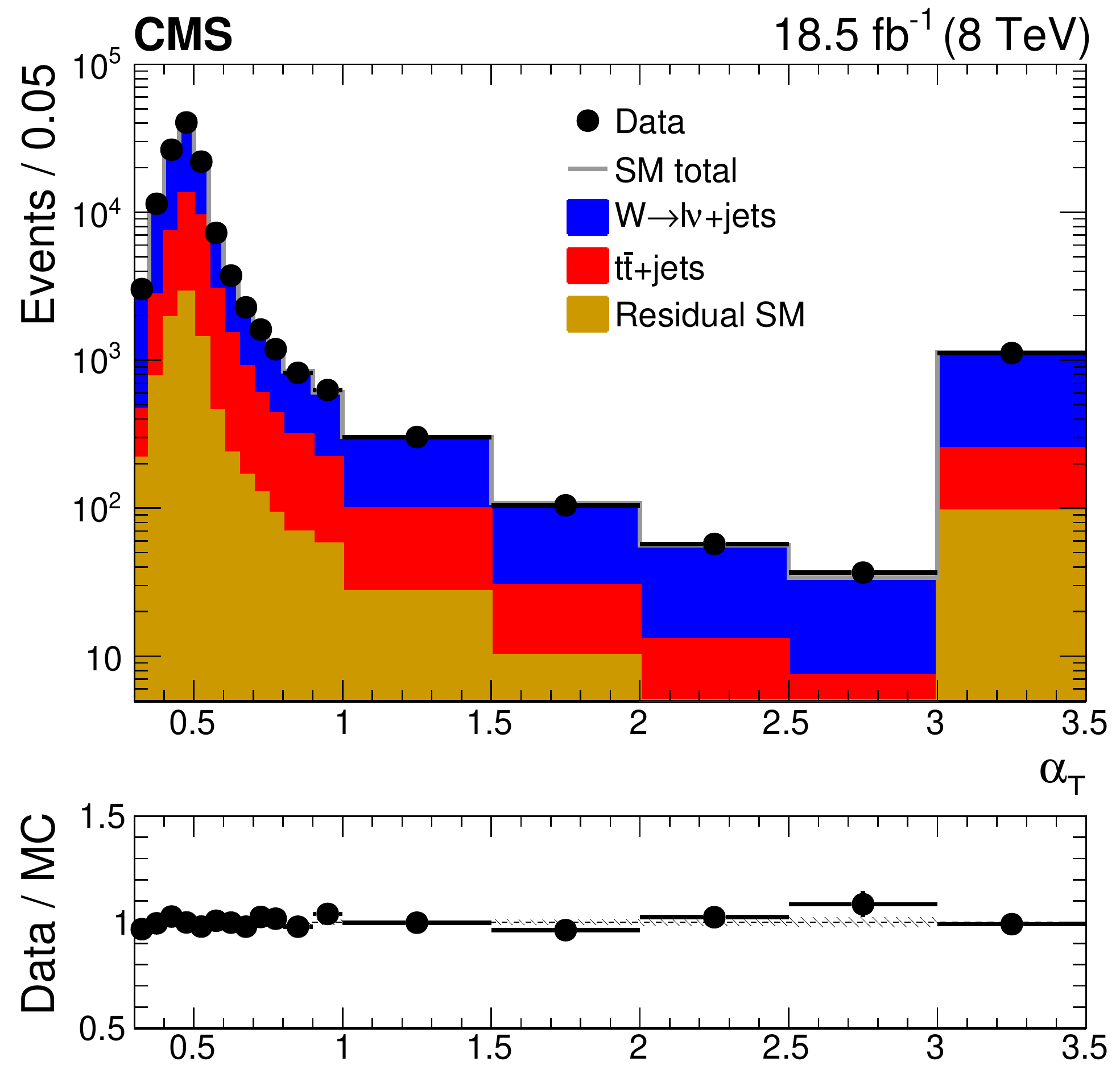} ~
  \includegraphics[width=0.49\textwidth]{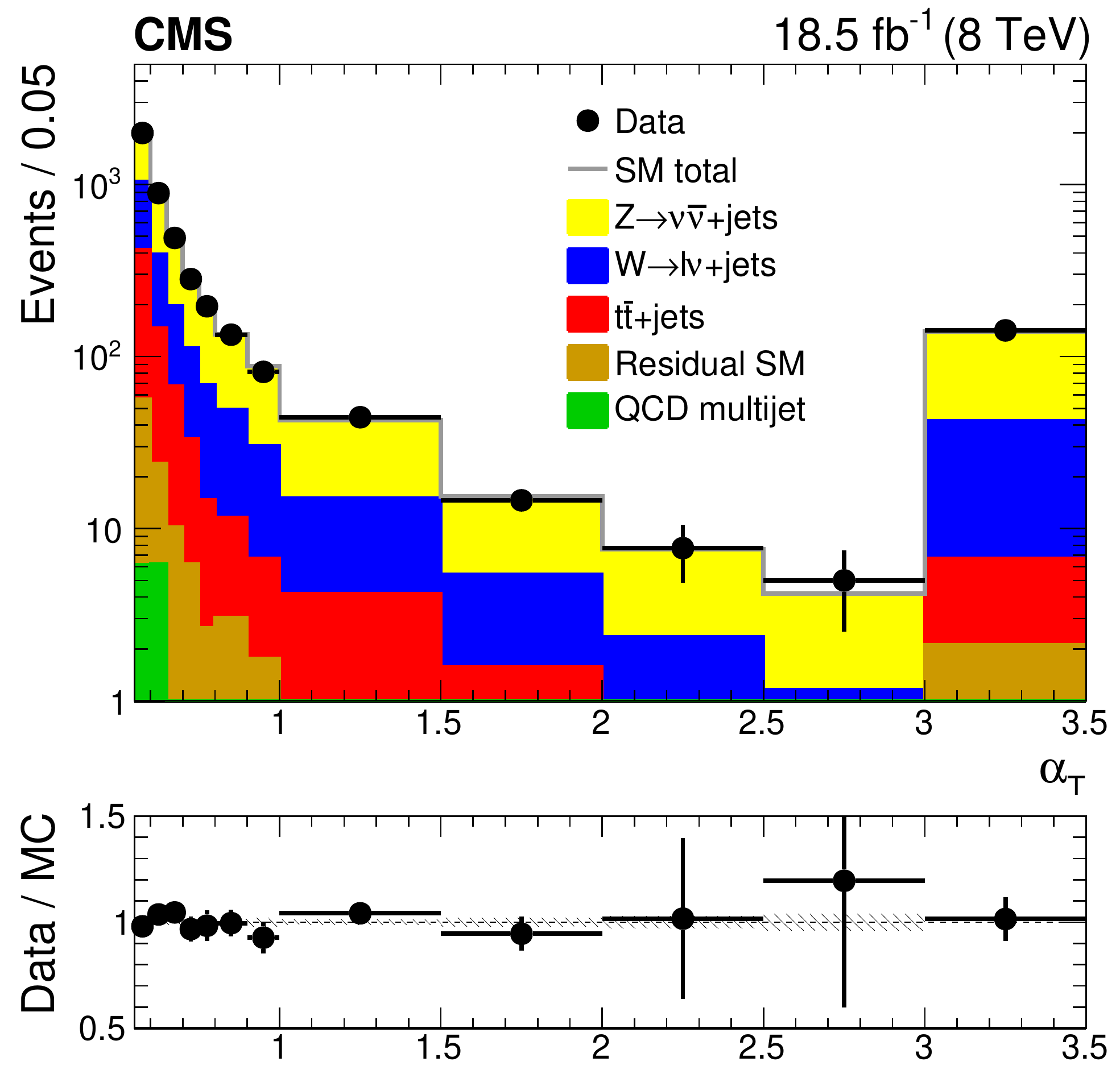}
  \caption{The \alphat distribution observed in data for event samples
    that are recorded with an inclusive set of trigger conditions and
    satisfy (\cmsLeft) the selection criteria that define the \mj
    control region or (\cmsRight) the criteria that define the signal
    region, with the additional requirement $\scalht > 375\GeV$. Event
    yields observed in data (solid circles) and SM expectations
    determined from simulation (solid histograms) are
    shown. Contributions from single top quark, diboson, Drell-Yan,
    and \ttbar + gauge boson production are collectively labelled
    ``Residual SM''. The final bin contains the overflow events. The
    lower panels show the ratios of the binned yields obtained from
    data and Monte Carlo (MC) simulation as a function of
    $\alpha_\text{T}$. The statistical uncertainties in the SM
    expectations are represented by the hatched
    areas. \label{fig:alphat} }
\end{figure}

The production of W and Z bosons in association with jets is simulated
with the leading-order (LO) \MADGRAPH 5.1.1.0~\cite{madgraph5} event
generator, with up to four additional partons considered in the matrix
element calculation. The production of \ttbar and single top quark
events is generated with the next-to-leading-order (NLO) \POWHEG
1.0~\cite{powheg, Nason:2004rx, Alioli:2010xd, Frixione:2007nw}
program. The LO \PYTHIA 6.4.26~\cite{pythia} program is used to
generate WW, WZ, and ZZ (diboson) events, and to describe parton
showering and hadronisation for all samples. The
{CTEQ6L1}~\cite{Pumplin:2002vw} and {CT10}~\cite{ct10} parton
distribution functions (PDFs) are used with \MADGRAPH and \POWHEG,
respectively. The description of the detector response is implemented
using the \GEANTfour~\cite{geant} package. The simulated samples are
normalised by the most accurate cross section calculations currently
available, usually up to next-to-next-to-leading-order (NNLO) accuracy
in QCD~\cite{xs-1, Gavin:2012sy, xs-2, xs-3, Czakon:2011xx}. To model
the effects of pileup, the simulated events are generated with a
nominal distribution of pp interactions per bunch crossing and then
reweighted to match the pileup distribution measured in data.

Figure~\ref{fig:alphat} shows the distributions of the \alphat
variable obtained from samples of events that satisfy the selection
criteria used to define the \mj control region and the signal
region. The inclusive requirements $\njet \geq 2$, $\nb \geq 0$, and
$\scalht > 200$ and 375\GeV for the two samples, respectively, are
imposed. The distributions illustrate the background composition of
the two samples as determined from simulation. While the figure also
demonstrates an adequate modelling of the \alphat variable with
simulated events, the method employed by the search to estimate the
nonmultijet backgrounds is designed to mitigate the effects of
simulation mismodelling.

The method relies on the use of transfer factors that are constructed
per bin, with a binning scheme defined identically to that of the
signal region in terms of \njet, \nb, and \scalht, for each control
sample in data. The transfer factors are determined using simulated
events, and are given by the ratios of the expected yields in the
corresponding bins of the signal region and control samples. The
transfer factors are used to extrapolate from the event yield measured
in a data control sample to the expectation for background from a
particular SM process or processes in the signal region. The method
aims to minimise the effects of simulation mismodelling, as many
systematic biases are expected to largely cancel in the ratios used to
define the transfer factors. Uncertainties in the transfer factors are
determined from a data-derived approach, described below.

The \mj data sample provides an estimate of the total contribution
from \ttbar and W boson production, as well as of the residual
contributions from single top quark, diboson, and Drell--Yan
($\PQq\PAQq \to \PZ/\gamma^* \to \ell^+\ell^-$) production. Two
independent estimates of the background from \znunujets events with
$\nb \leq 1$ are determined, one from the \gj data sample and the
other from the \mmj data sample, which are considered simultaneously
in the likelihood function described in Section~\ref{sec:results}. The
\gj and \zmumujets processes have similar kinematic properties when
the photon or muons are ignored in the determination of \ETmiss and
\HTmiss~\cite{Bern:2011pa}, although the acceptances differ. An
advantage of the \gj process is its much larger production cross
section compared to the \znunujets process.

In the case of events with $\nb \geq 2$, the \mj sample is also used
to estimate the small \znunujets background because of the limited
event counts in the \mmj and \gj control samples.  The method relies
on the use of \wmujets events to predict the \znunujets
background~\cite{RA1Paper, RA1Paper2011FULL, RA1Paper2012}. The method
corrects for \ttbar contamination in the \mj sample, which can be
significant in the presence of jets identified as originating from b
quarks. However, while the \ttbar contamination increases with
increasing \nb, the \znunujets background is reduced to a subdominant
level relative to other backgrounds. The method is validated in data
control regions defined by samples of events categorised according to
\nb. In summary, only the \mj sample is used to estimate the total SM
background for events with $\nb \geq 2$, whereas all three data
control samples are used for events with $\nb \leq 1$.

To maximise sensitivity to new-physics signatures with a large number
of b quarks, a method is employed that allows event yields for a given
b quark jet multiplicity to be predicted with a higher statistical
precision than obtained directly from simulation, particularly for
events with a large number of b quark jets ($N_\cPqb \ge
2$)~\cite{RA1Paper2012}. The method relies on generator-level
information contained in the simulation to determine the distribution
of \nb for a sample of events categorised according to \njet and
\scalht. First, simulated events are categorised according to the
number of jets per event that are matched to underlying b quarks
($N_\cPqb^\text{gen}$), c quarks ($N_\cPqc^\text{gen}$), and
light-flavoured quarks or gluons ($N_\cPq^\text{gen}$). Second, the
efficiency $\epsilon$ with which b quark jets are identified, and the
misidentification probabilities for c quarks and light-flavour
partons, $f_\cPqc$ and $f_\cPq$, respectively, are also determined
from simulation, with each quantity averaged over jet \pt and $\eta$
per event category. Corrections to $\epsilon$, $f_\cPqc$, and $f_\cPq$
are applied on a jet-by-jet basis as a function of \pt and $\eta$ so
that they match the corresponding quantity measured in
data~\cite{Chatrchyan:2012jua}. Finally, $N_{\cPqb}^{\text{tag}}$,
$N_{\cPqc}^{\text{tag}}$, and $N_{\cPq}^{\text{tag}}$ are,
respectively, the number of jets identified (``tagged'') as
originating from b quarks per event when the underlying parton is a b
quark, c quark, or a light-flavoured quark or gluon, and
$P(N_{\cPqb}^{\text{tag}} ; N_{\cPqb}^{\text{gen}}, \epsilon)$,
$P(N_{\cPqc}^{\text{tag}} ; N_{\cPqc}^{\text{gen}}, f_\cPqc)$, and
$P(N_{\cPq}^{\text{tag}} ; N_{\cPq}^{\text{gen}}, f_\cPq)$ are the
binomial probabilities for this to happen. These quantities are
sufficient to estimate how events are distributed according to
$N_\cPqb$ per (\njet, \scalht) category when summing over all relevant
combinations that satisfy the requirements $\njet = N_\cPqb^\text{gen}
+ N_\cPqc^\text{gen} + N_\cPq^\text{gen}$ and $\nb =
N_{\cPqb}^{\text{tag}} + N_{\cPqc}^{\text{tag}} +
N_{\cPq}^{\text{tag}}$.

The event yields determined with the method described above are
subsequently used to determine the transfer factors binned according
to \nb (in addition to \njet and \scalht). The uncertainties in the
transfer factors obtained from simulation are evaluated through sets
of closure tests based on events from the data control
regions~\cite{RA1Paper2012}. Each set uses the observed event counts
in up to eleven bins in \scalht for a given sample of events, along
with the corresponding (\scalht-dependent) transfer factors obtained
from simulation, to determine \scalht-dependent predictions
$N_\text{pred}(\scalht)$ for yields in another event sample. The two
samples are taken from different data control regions, or are subsets
of the same data control sample with differing requirements on \njet
or \nb. The predictions $N_\text{pred}(\scalht)$ are compared with the
\scalht-binned observed yields $N_\text{obs}(\scalht)$ and the level
of closure is defined by the deviation of the ratio $(N_\text{obs} -
N_\text{pred})/N_\text{pred}$ from zero. A large number of tests are
performed to probe key aspects of the modelling that may introduce an
\njet- or \scalht-dependent source of bias in the transfer
factors~\cite{RA1Paper2012}.

\begin{figure}[!htbp]
  \centering
      \includegraphics[width=\cmsFigWidth]{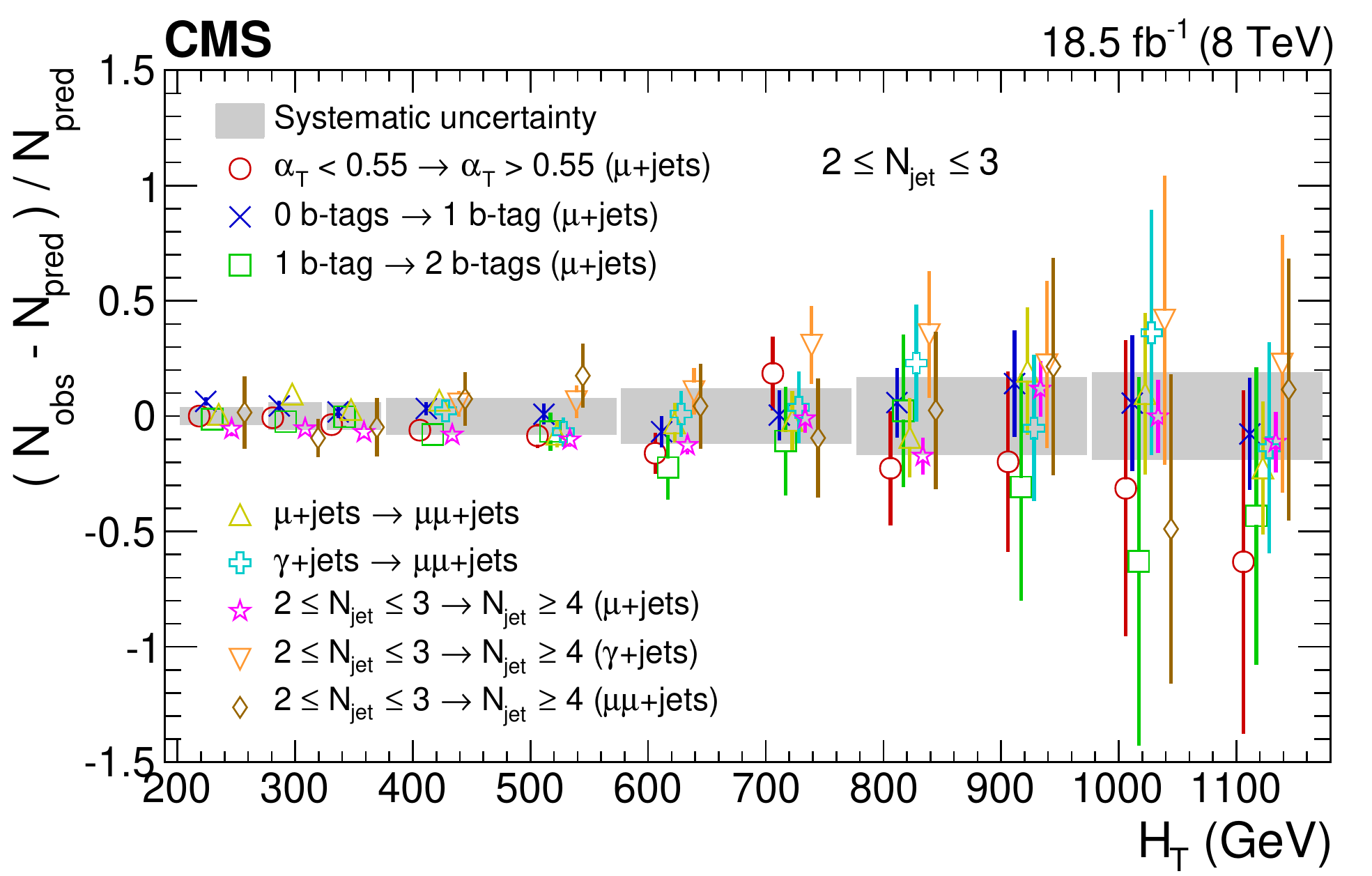}
      \includegraphics[width=\cmsFigWidth]{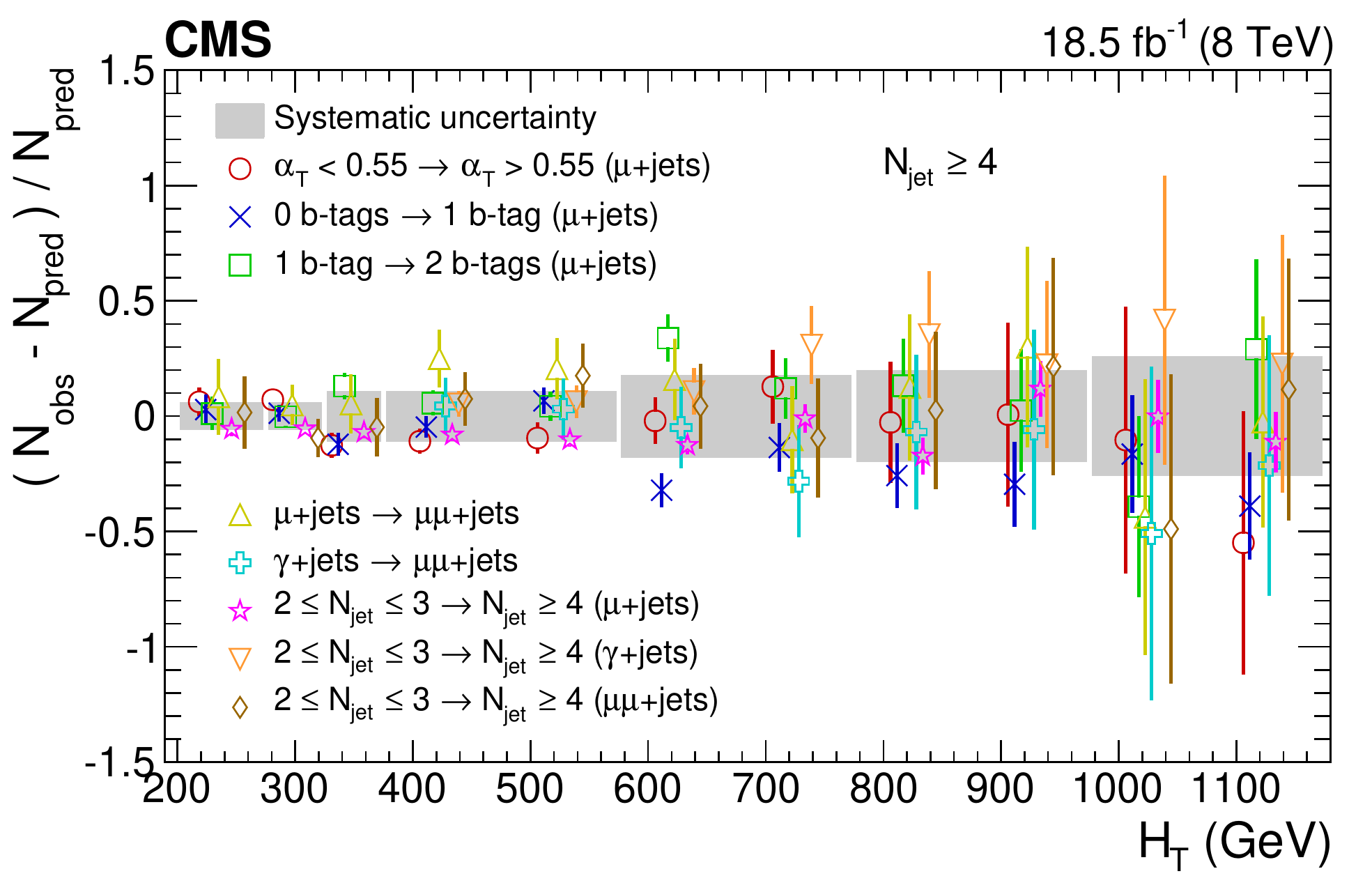}
    \caption{
      Ratio $(N_\text{obs} - N_\text{pred})/N_\text{pred}$ as a
      function of \scalht for different event categories and/or
      control regions for (upper) events with two or three jets, and
      (lower) events with four or more jets; ``b tag'' refers to a
      reconstructed b quark candidate.
      Error bars represent statistical uncertainties only, while the
      grey shaded bands represent the \njet- and \scalht-dependent
      uncertainties assumed in the transfer factors, as determined
      from the procedure described in the text.
      \label{fig:closure}
    }
\end{figure}

Systematic uncertainties are determined from core sets of closure
tests, of which the results are shown in Fig.~\ref{fig:closure}. Five
sets of tests are performed independently for each of the two \njet
categories, and a further three sets that are common to both \njet
categories. The tests aim to probe for the presence of statistically
significant biases that could arise due to limitations in the
method. For each \njet category, the first three sets of closure tests
are performed using the \mj sample. The first set probes the modelling
of the \alphat distribution for events containing genuine \ptvecmiss
from neutrinos (open circle markers). Two sets (crosses, squares)
probe the relative composition between \wjets and top events and the
modelling of the reconstruction of b quark jets. The fourth set
(triangles) validates the modelling of vector boson production by
connecting the \mj and \mmj control samples, which are enriched in
\wjets and \zjets events, respectively. The fifth set (swiss crosses)
deals with the consistency between the \gj and \mmj samples, which are
both used to provide an estimate of the \znunujets background. Three
further sets of closure tests (stars, inverted triangles, diamonds),
one per data control sample, probe the simulation modelling of the
\njet distribution for a range of background compositions.

\begin{table*}[!thb]
  \topcaption{Systematic uncertainties (\%) in the transfer factors,
    in intervals of \njet and \scalht.}
  \label{tab:syst-values}
  \centering
  \begin{tabular}{ l*{7}r }
    \hline
            & \multicolumn{7}{c}{\scalht region (\GeVns)}                                \\
    \cline{2-8}
    \njet   & \multicolumn{1}{c}{200--275} & \multicolumn{1}{c}{275--325} & \multicolumn{1}{c}{325--375} & \multicolumn{1}{c}{375--575} & \multicolumn{1}{c}{575--775} & \multicolumn{1}{c}{775-975} & \multicolumn{1}{c}{$>975$} \\
    \hline
    2--3    & 4        & 6        & 6        & 8        & 12       & 17      & 19     \\
    $\geq$4 & 6        & 6        & 11       & 11       & 18       & 20      & 26     \\
    \hline
  \end{tabular}
\end{table*}

The closure tests reveal no significant biases or dependency on \njet
nor \scalht. Systematic uncertainties in the transfer factors are
determined from the variance in $(N_\text{obs} -
N_\text{pred})/N_\text{pred}$, weighted to account for statistical
uncertainties, for all closure tests within an individual \scalht bin
in the range $200 < \scalht < 375\GeV$ and for each \njet
category. For the region $\scalht > 375\GeV$, all tests within
200\GeV-wide intervals in \scalht, defined by pairs of adjacent bins,
are combined to determine the systematic uncertainty, which is assumed
to be fully correlated for bins within each interval, and fully
uncorrelated for different \scalht intervals and \njet categories. The
magnitudes of the systematic uncertainties are indicated by shaded
grey bands in Fig.~\ref{fig:closure} and summarised in
Table~\ref{tab:syst-values}. The same (uncorrelated) value of
systematic uncertainty is assumed for each \nb category.
An independent study is performed to assess the effect of
uncertainties in the simulation modelling of the efficiency and
misidentification rates for jets originating from b quarks and from
light-flavoured quarks or gluons. These uncertainties are found to be
at the sub-percent level, subdominant relative to the values in
Table~\ref{tab:syst-values}, and therefore considered to be
negligible.

\section{Results and interpretation\label{sec:results}}

For a given category of events satisfying requirements on both \njet
and \nb, a likelihood model of the observations in all data samples is
used to obtain a consistent prediction of the SM backgrounds and to
test for the presence of a variety of signal models. This is written
as:
\begin{equation}
  \label{likelihood}
  \begin{aligned}
  L_{\njet,\,\nb}  & =  L_\mathrm{SR} L_{\mu} L_{\mu\mu} L_{\gamma}, & (0 \leq \nb \leq 1) \\
  L_{\njet,\,\nb}  & =  L_\mathrm{SR} L_{\mu}, & (\nb \geq 2)
  \end{aligned}
\end{equation}
where $L_\mathrm{SR} = \prod_{i} \mathrm{Pois}(
n^{i} \, | \, b^{i} +
s^{i} )$ is a likelihood function comprising a
product of Poisson terms that describe the yields in each of the
\scalht bins of the signal region for given values of \njet and \nb.
In each bin of \scalht (index $i$), the observation
$n^{i}$ is modelled as a Poisson variable
distributed about the sum of the SM expectation
$b^{i}$ and a potential contribution from a signal
model $s^{i}$ (assumed to be zero in the following
discussion). The contribution from multijet production is assumed to
be zero, based on the studies described in Section~\ref{sec:multijet}.
The SM expectations in the signal region are related to the expected
yields in the \mj, \mmj, and \gj control samples via the transfer
factors derived from simulation. Analogous to $L_\mathrm{SR}$, the
likelihood functions $L_\mu$, $L_{\mu\mu}$, and $L_\gamma$ describe
the yields in the \scalht bins of the \mj, \mmj, and \gj control
samples for the same values of \njet and \nb as the signal region.
For the category of events with $\nb \geq 2$, only the \mj control
sample is used in the likelihood to determine the total contribution
from all nonmultijet SM backgrounds in the signal region. The
systematic uncertainties in the transfer factors, determined from the
ensemble of closure tests described above and with magnitudes in the
range 4--26\% (Table~\ref{tab:syst-values}), are accommodated in the
likelihood function through a nuisance parameter associated with each
transfer factor used in the background estimation for each (\njet,
\nb) category and \scalht interval. The \scalht intervals are defined
by pairs of adjacent \scalht bins for the region $\scalht > 375\GeV$,
as described in Section~\ref{sec:ewk}, and so adjacent bins share the
same nuisance parameter. The measurements of these parameters are
assumed to follow a lognormal distribution.

\begin{table*}[!t]
  \topcaption{
    Observed event yields in data and the ``a priori'' SM expectations
    determined from event counts in the data control samples and
    transfer factors from simulation, in bins of \scalht, and
    categorised according to \njet and \nb. Also shown are the SM
    expectations (labelled ``SM'') obtained from a combined fit to
    control and signal regions under the SM hypothesis. The quoted
    uncertainties include the statistical as well as systematic
    components. For each row that lists fewer than the full set of
    columns, the final entry represents values obtained for an open
    final \scalht bin.
  }
  \label{tab:fit-result}
  \centering
  \resizebox{\textwidth}{!}{
  \renewcommand*{\arraystretch}{1.4}
  \begin{tabular}{ lllllllllllll }
    \hline
    Category &  & \multicolumn{11}{c}{\scalht (\GeVns)} \\
    \cline{3-13}
    (\njet,\,\nb)
             &
             & 200--275
             & 275--325
             & 325--375
             & 375--475
             & 475--575
             & 575--675
             & 675--775
             & 775--875
             & 875--975
             & 975--1075
             & 1075--$\infty$                           \\
    \hline
    (2--3,\,0)
             & Data
             & $13090$
             & $5331$
             & $3354$
             & $2326$
             & $671$
             & $206$
             & $76$
             & $29$
             & $10$
             & $9$
             & $2$                                      \\
    (2--3,\,0)
             & a priori
             & $12410^{+370}_{-410}$
             & $5540^{+340}_{-230}$
             & $3330^{+130}_{-170}$
             & $2400^{+120}_{-90}$
             & $663^{+34}_{-26}$
             & $225^{+21}_{-17}$
             & $68.5^{+6.9}_{-6.7}$
             & $26.5^{+3.9}_{-3.0}$
             & $10.3^{+1.9}_{-2.1}$
             & $5.1^{+1.0}_{-1.1}$
             & $4.5^{+0.9}_{-0.9}$                      \\
    (2--3,\,0)
             & SM
             & $13030^{+90}_{-120}$
             & $5348^{+85}_{-67}$
             & $3351^{+56}_{-50}$
             & $2351^{+38}_{-45}$
             & $655^{+14}_{-11}$
             & $218^{+12}_{-17}$
             & $68.5^{+4.9}_{-4.8}$
             & $27.2^{+3.0}_{-3.0}$
             & $10.4^{+1.5}_{-1.6}$
             & $5.6^{+1.0}_{-1.0}$
             & $4.3^{+0.7}_{-1.0}$                      \\\\[-2ex]
    (2--3,\,1)
             & Data
             & $1733$
             & $833$
             & $527$
             & $356$
             & $90$
             & $31$
             & $6$
             & $4$
             & $1$
             & $0$
             & $1$                                      \\
    (2--3,\,1)
             & a priori
             & $1669^{+65}_{-67}$
             & $853^{+50}_{-46}$
             & $525^{+37}_{-24}$
             & $391^{+23}_{-21}$
             & $94.3^{+6.0}_{-5.6}$
             & $24.5^{+2.5}_{-3.6}$
             & $9.0^{+1.2}_{-1.4}$
             & $2.8^{+0.6}_{-0.8}$
             & $2.5^{+0.8}_{-0.9}$
             & $0.3^{+0.2}_{-0.1}$
             & $0.2^{+0.1}_{-0.1}$                      \\
    (2--3,\,1)
             & SM
             & $1711^{+37}_{-33}$
             & $839^{+21}_{-25}$
             & $526^{+20}_{-17}$
             & $372^{+12}_{-14}$
             & $90.6^{+5.1}_{-4.6}$
             & $25.8^{+2.9}_{-2.6}$
             & $8.7^{+0.8}_{-1.4}$
             & $3.0^{+0.7}_{-0.6}$
             & $2.2^{+0.8}_{-0.6}$
             & $0.3^{+0.2}_{-0.1}$
             & $0.2^{+0.1}_{-0.2}$                      \\\\[-2ex]
    (2--3,\,2)
             & Data
             & $172$
             & $116$
             & $101$
             & $55$
             & $16$
             & $9$
             & $0$
             & $0$
             & $0$                                      \\
    (2--3,\,2)
             & a priori
             & $187^{+7}_{-8}$
             & $118^{+7}_{-7}$
             & $98.7^{+7.1}_{-7.0}$
             & $61.3^{+5.9}_{-5.5}$
             & $12.3^{+1.7}_{-1.0}$
             & $2.8^{+0.5}_{-0.6}$
             & $0.7^{+0.2}_{-0.2}$
             & $0.2^{+0.1}_{-0.1}$
             & $<$0.1                                   \\
    (2--3,\,2)
             & SM
             & $184^{+5}_{-7}$
             & $117^{+7}_{-5}$
             & $99.4^{+5.4}_{-4.6}$
             & $60.2^{+3.5}_{-3.8}$
             & $12.4^{+1.2}_{-1.0}$
             & $3.3^{+0.6}_{-0.5}$
             & $0.7^{+0.2}_{-0.2}$
             & $0.2^{+0.1}_{-0.1}$
             & $<$0.1                                   \\\\[-2ex]
    ($\geq$4,\,0)
             & Data
             & $99$
             & $568$
             & $408$
             & $336$
             & $211$
             & $117$
             & $38$
             & $13$
             & $9$
             & $4$
             & $6$                                      \\
    ($\geq$4,\,0)
             & a priori
             & $108^{+10}_{-12}$
             & $497^{+34}_{-36}$
             & $403^{+36}_{-33}$
             & $327^{+25}_{-22}$
             & $193^{+14}_{-13}$
             & $95^{+13}_{-11}$
             & $40.3^{+5.9}_{-4.4}$
             & $14.5^{+3.5}_{-2.4}$
             & $7.1^{+1.7}_{-1.4}$
             & $3.2^{+0.7}_{-1.0}$
             & $2.9^{+0.7}_{-0.5}$                      \\
    ($\geq$4,\,0)
             & SM
             & $104^{+6}_{-8}$
             & $544^{+21}_{-18}$
             & $407^{+18}_{-18}$
             & $337^{+15}_{-10}$
             & $202^{+10}_{-8}$
             & $105^{+9}_{-7}$
             & $42.5^{+4.5}_{-3.3}$
             & $14.3^{+1.7}_{-2.5}$
             & $7.5^{+1.4}_{-1.5}$
             & $3.5^{+0.8}_{-0.8}$
             & $3.4^{+1.0}_{-0.7}$                      \\\\[-2ex]
    ($\geq$4,\,1)
             & Data
             & $38$
             & $195$
             & $210$
             & $159$
             & $83$
             & $33$
             & $7$
             & $10$
             & $4$
             & $1$
             & $1$                                      \\
    ($\geq$4,\,1)
             & a priori
             & $39.2^{+3.0}_{-3.5}$
             & $215^{+12}_{-16}$
             & $208^{+24}_{-22}$
             & $150^{+15}_{-11}$
             & $75.8^{+7.8}_{-6.6}$
             & $28.6^{+3.8}_{-3.7}$
             & $10.3^{+2.1}_{-1.4}$
             & $5.1^{+1.3}_{-0.9}$
             & $2.0^{+0.7}_{-0.5}$
             & $0.8^{+0.4}_{-0.3}$
             & $0.9^{+0.6}_{-0.4}$                      \\
    ($\geq$4,\,1)
             & SM
             & $38.9^{+2.2}_{-3.7}$
             & $206^{+12}_{-10}$
             & $209^{+13}_{-10}$
             & $157^{+9}_{-9}$
             & $79.3^{+5.2}_{-4.7}$
             & $29.4^{+3.8}_{-2.2}$
             & $9.9^{+1.9}_{-1.3}$
             & $6.2^{+1.2}_{-1.1}$
             & $2.3^{+0.7}_{-0.7}$
             & $0.9^{+0.3}_{-0.3}$
             & $0.9^{+0.3}_{-0.4}$                      \\\\[-2ex]
    ($\geq$4,\,2)
             & Data
             & $16$
             & $81$
             & $88$
             & $64$
             & $43$
             & $14$
             & $5$
             & $1$
             & $1$                                      \\
    ($\geq$4,\,2)
             & a priori
             & $12.3^{+1.0}_{-1.0}$
             & $76.7^{+5.6}_{-5.2}$
             & $93^{+11}_{-9}$
             & $63.0^{+7.8}_{-5.7}$
             & $34.0^{+3.6}_{-3.4}$
             & $10.1^{+2.6}_{-1.8}$
             & $3.4^{+0.9}_{-0.6}$
             & $1.0^{+0.2}_{-0.2}$
             & $0.7^{+0.1}_{-0.2}$                      \\
    ($\geq$4,\,2)
             & SM
             & $12.5^{+1.0}_{-1.0}$
             & $77.8^{+4.7}_{-4.6}$
             & $90.2^{+9.0}_{-6.5}$
             & $66.1^{+4.6}_{-4.8}$
             & $36.3^{+3.4}_{-2.9}$
             & $11.4^{+1.8}_{-1.9}$
             & $3.9^{+0.8}_{-0.7}$
             & $1.0^{+0.2}_{-0.3}$
             & $0.7^{+0.1}_{-0.2}$                      \\\\[-2ex]
    ($\geq$4,\,3)
             & Data
             & $0$
             & $7$
             & $5$
             & $5$
             & $6$
             & $1$
             & $1$
             & $0$
             & $0$                                      \\
    ($\geq$4,\,3)
             & a priori
             & $1.1^{+0.2}_{-0.1}$
             & $8.2^{+0.6}_{-0.9}$
             & $11.1^{+2.0}_{-1.6}$
             & $7.4^{+1.1}_{-1.0}$
             & $4.0^{+0.5}_{-0.6}$
             & $1.1^{+0.3}_{-0.3}$
             & $0.4^{+0.2}_{-0.1}$
             & $0.1^{+0.1}_{-0.0}$
             & $<$0.1                                   \\
    ($\geq$4,\,3)
             & SM
             & $1.1^{+0.2}_{-0.2}$
             & $8.1^{+0.9}_{-0.9}$
             & $9.9^{+1.5}_{-1.3}$
             & $7.2^{+0.9}_{-0.7}$
             & $4.1^{+0.6}_{-0.6}$
             & $1.1^{+0.3}_{-0.3}$
             & $0.4^{+0.1}_{-0.1}$
             & $0.1^{+0.1}_{-0.0}$
             & $<$0.1                                   \\\\[-2ex]
    ($\geq$4,\,$\geq$4)
             & Data
             & $0$
             & $0$
             & $0$
             & $2$                                      \\
    ($\geq$4,\,$\geq$4)
             & a priori
             & $<$0.1
             & $0.2^{+0.1}_{-0.1}$
             & $0.5^{+0.3}_{-0.3}$
             & $0.3^{+0.2}_{-0.2}$                      \\
    ($\geq$4,\,$\geq$4)
             & SM
             & $<$0.1
             & $0.1^{+0.1}_{-0.1}$
             & $0.4^{+0.2}_{-0.3}$
             & $0.4^{+0.2}_{-0.2}$                      \\

    \hline
  \end{tabular}
  }
\end{table*}

Table~\ref{tab:fit-result} summarises the observed event yields and
expected number of events from SM processes in the signal region as a
function of \njet, \nb, and \scalht. The ``a priori'' SM expectations
are determined from event counts in the data control samples and
transfer factors from simulation, and are therefore independent of the
signal region. No significant discrepancies are observed between the
``a priori'' SM expectations and the observed event yields. In
addition, a simultaneous fit to data in the signal region and in up to
three control regions is performed. The likelihood function is
maximised over all fit parameters under the SM-only hypothesis in
order to estimate the yields from SM processes in each bin in all
regions, in the absence of an assumed contribution from signal
events. Table~\ref{tab:fit-result} summarises these estimates
(labelled ``SM'') for the signal region.
A goodness-of-fit test is performed to quantify the degree of
compatibility between the observed yields and the expectations under
the background-only hypothesis. The test is based on a log likelihood
ratio and the alternative hypothesis is defined by a ``saturated''
model~\cite{sat-llk}. The $p$-value probabilities for all \njet and
\nb categories are found to be uniformly distributed, with a minimum
value of 0.19.

The results of this search are interpreted in terms of limits on the
parent sparticle and LSP masses in the parameter space of simplified
models~\cite{Alwall:2008ag, Alwall:2008va, sms} that represent the
direct pair production of top squarks and the decay modes $\PSQt
\to\PQc\PSGczDo$, $\PSQt \to{\PQb \ffbp} \PSGczDo$, $\PSQt \to \PQb
\PSGcpmDo$ followed by $\PSGcpmDo \to \PW^\pm \PSGczDo$, and $\PSQt
\to \PQt \PSGczDo$. The \cls method~\cite{read,junk} is used to
determine upper limits at the 95\% confidence level (CL) on the
production cross section of a signal model, using the one-sided
(LHC-style) profile likelihood ratio as the test
statistic~\cite{higgs-comb}. The sampling distributions for the test
statistic are generated from pseudo-experiments using the respective
maximum likelihood values of nuisance parameters determined from a
simultaneous fit to the pseudo-data, in the 75 bins of the signal
region and in the corresponding bins of up to three control samples,
under the SM background-only and signal + background hypotheses. The
potential contributions of signal events to each of the signal and
control samples are considered, but the only significant contribution
occurs in the signal region and not the control samples.

The event samples for the simplified models are generated with the LO
\MADGRAPH 5.1.1.0 generator, which considers up to two additional
partons in the matrix element calculation. Inclusive,
process-dependent, NLO calculations of SUSY production cross sections,
with next-to-leading-logarithmic (NLL) corrections, are obtained with
the program \PROSPINO 2.1~\cite{Beenakker:1996ch, PhysRevD.80.095004,
  PhysRevLett.102.111802, PhysRevD.80.095004, 1126-6708-2009-12-041,
  doi:10.1142/S0217751X11053560, susy-nlo-nll}. All events are
generated using the \textsc{CTEQ6L1} PDFs. As for SM processes, the
simulated events are generated with a nominal pileup distribution and
then reweighted to match the distribution observed in data. The
detector response is provided by the CMS fast simulation
package~\cite{fastsim}.

Experimental uncertainties in the expected signal yields are
considered. Contributions to the overall systematic uncertainty arise
from various sources such as the uncertainties from the choice of
PDFs, the jet energy scale, the modelling of the efficiency and
misidentification probability of b quark jets in simulation, the
integrated luminosity~\cite{lumi}, and various event selection
criteria. The magnitude of each contribution depends on the model, the
masses of the parent sparticle and LSP, and the event category under
consideration. Uncertainties in the jet energy scale are typically
dominant ($\sim$15\%) for models with mass splittings that satisfy
$\dm > m_\PQt$, where $m_\PQt$ is the top quark mass. The acceptance
for models with mass splittings satisfying $\dm < m_\PQt$ is due in
large part to ISR, the modelling of which contributes the dominant
systematic uncertainty for systems with a compressed mass spectrum. An
uncertainty of $\sim$20\% is determined by comparing the simulated and
measured \pt spectra of the system recoiling against the ISR jets in
\ttbar events, using the technique described in
Ref.~\cite{single-lepton-stop}. For the aforementioned simplified
models, the effect of uncertainties in the distribution of signal
events is generally small compared with the uncertainties in the
experimental acceptance. The total systematic uncertainty in the yield
of signal is found to be in the range 5--36\%, depending on \njet and
\nb, and is taken into account through a nuisance parameter that
follows a lognormal distribution.

\begin{figure*}[!tbhp]
\centering
      \includegraphics[width=0.40\textwidth]{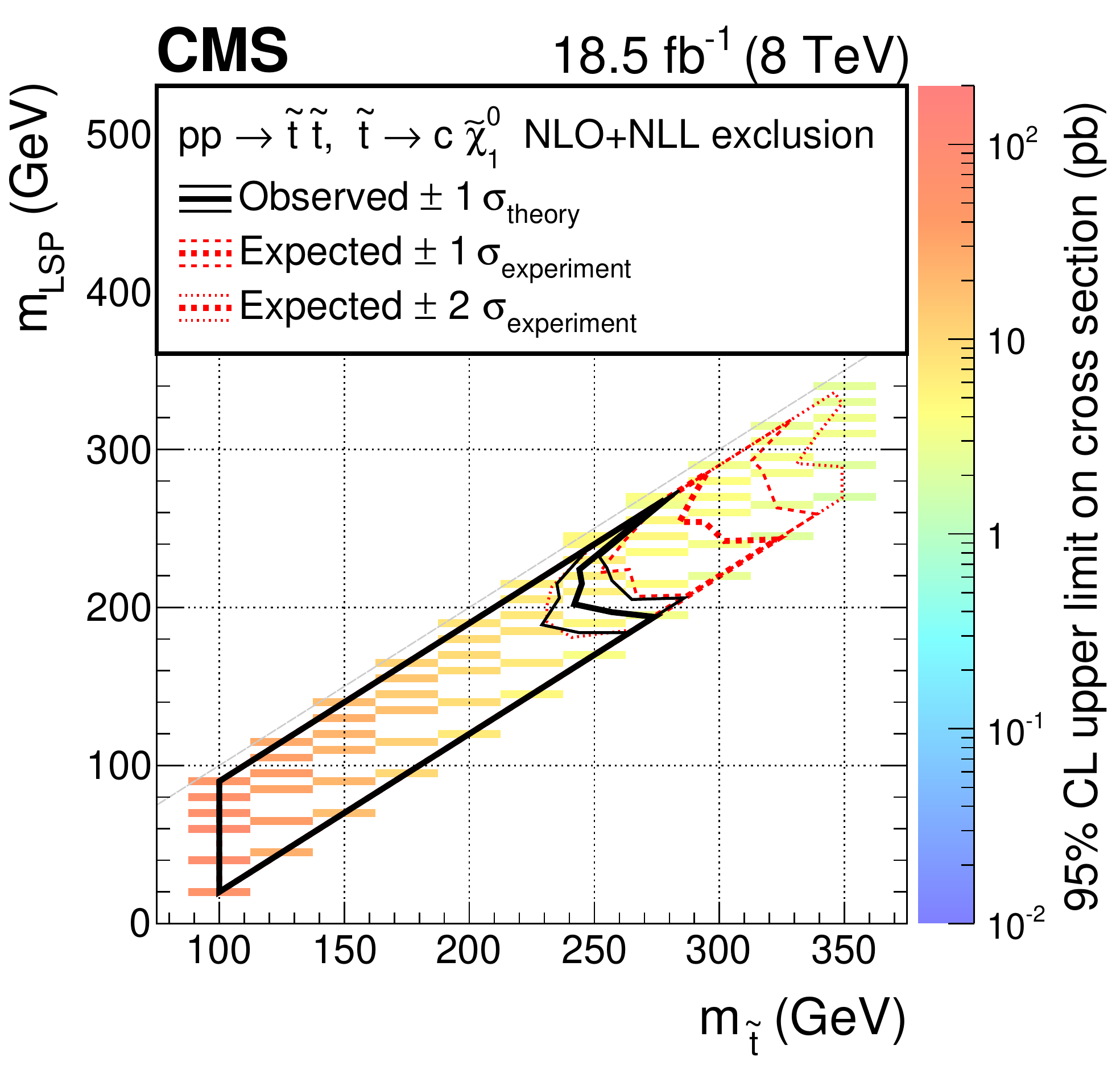}
      \includegraphics[width=0.40\textwidth]{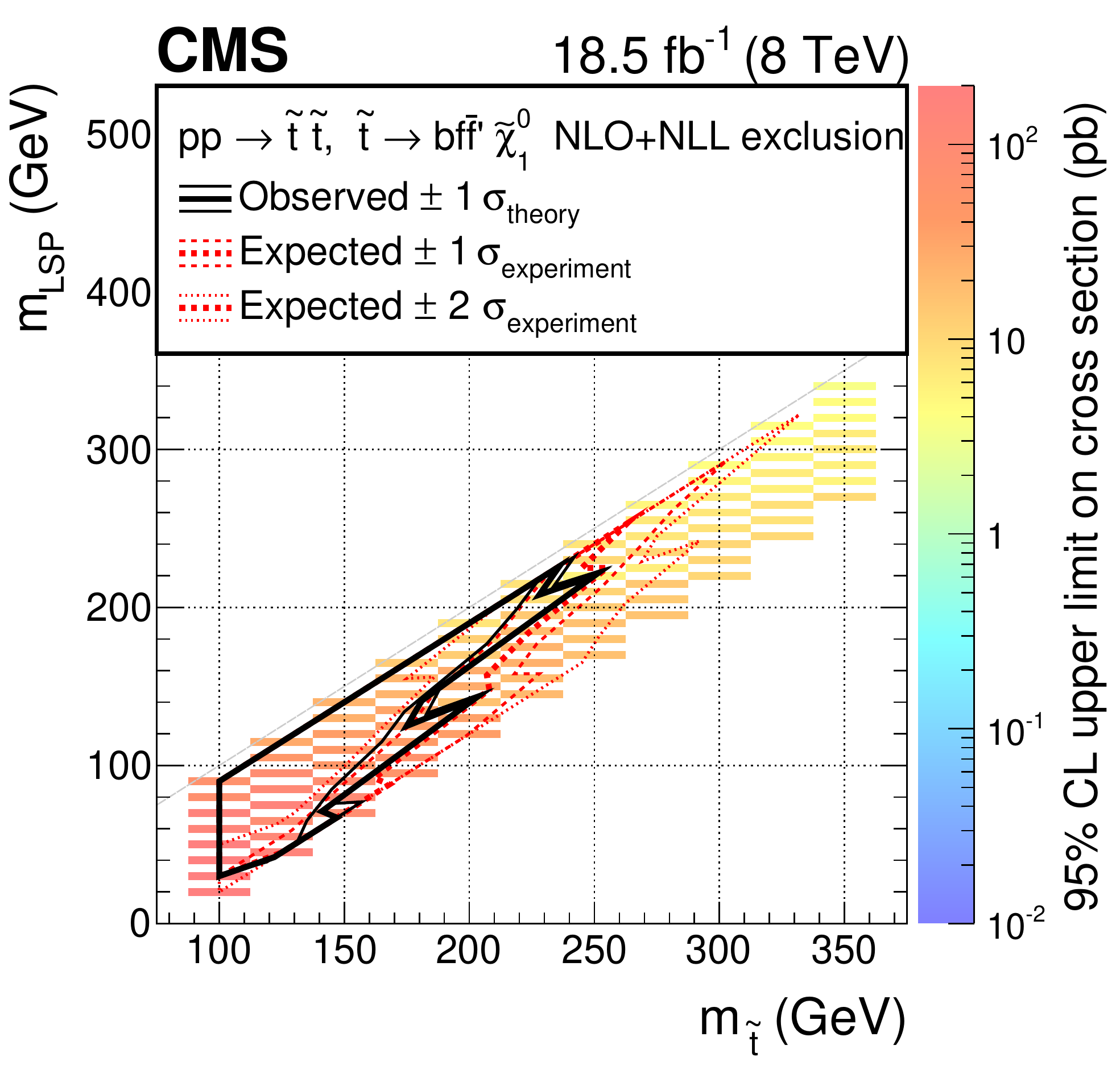}
      \includegraphics[width=0.40\textwidth]{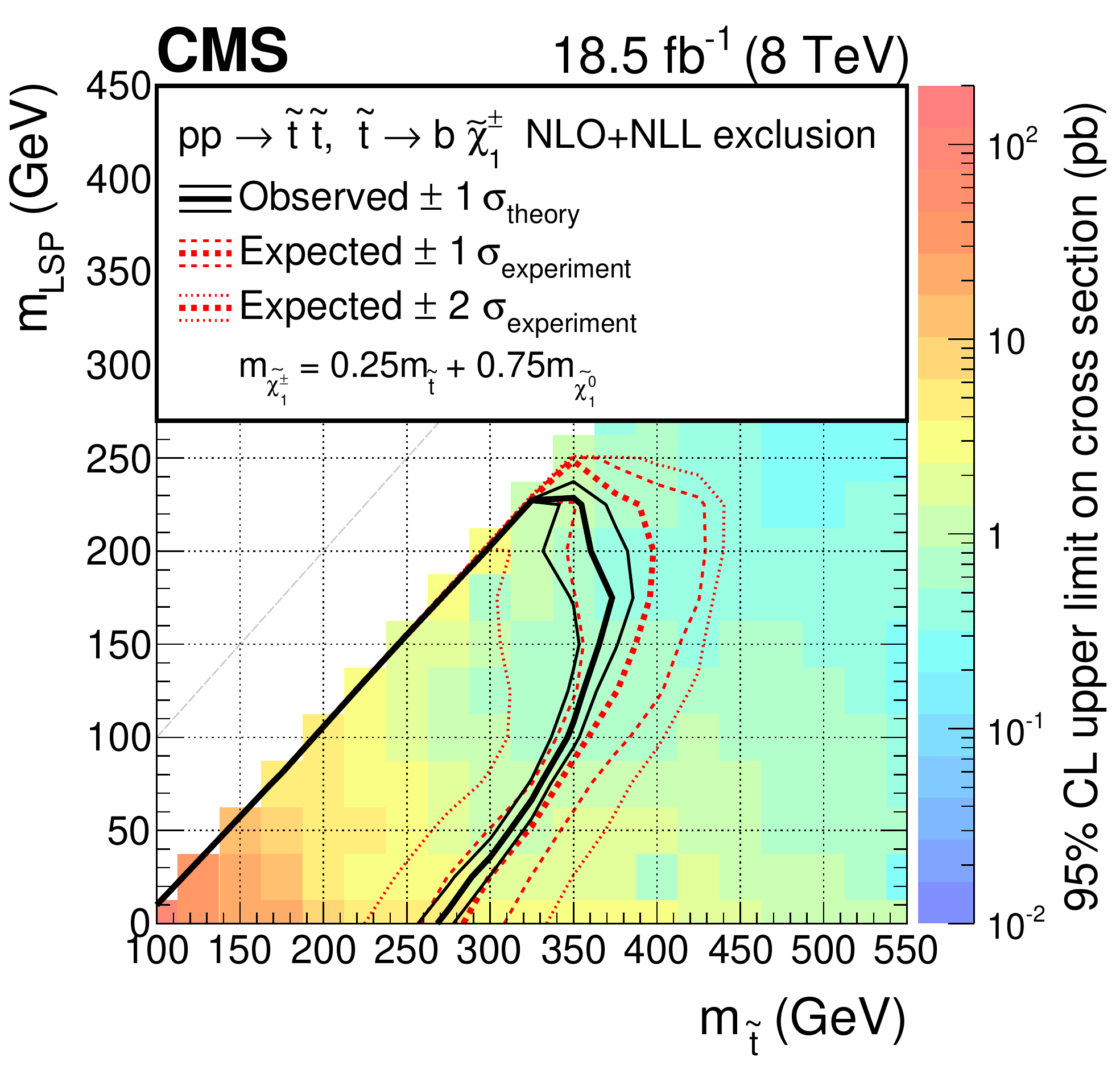}
      \includegraphics[width=0.40\textwidth]{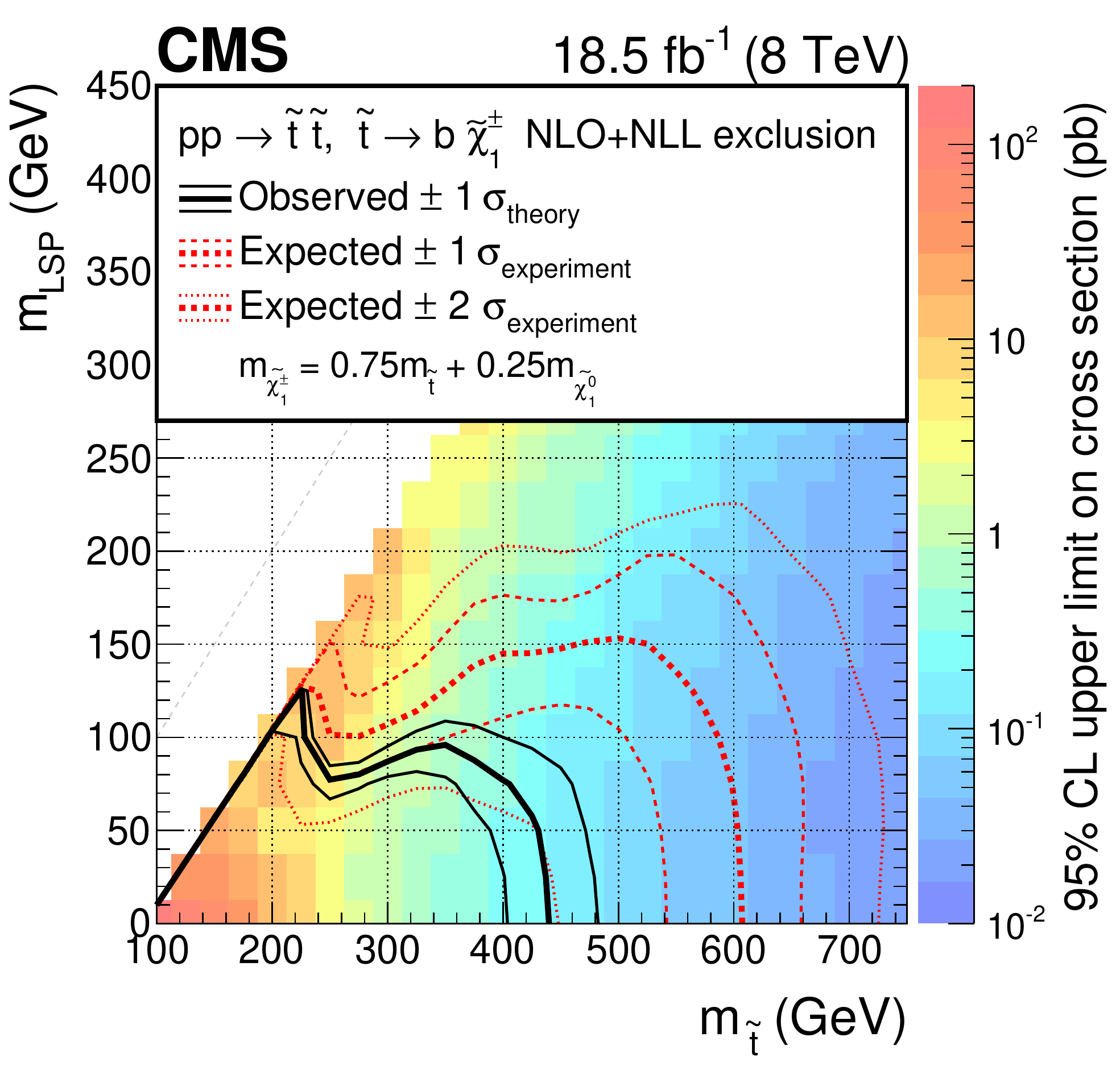}
      \includegraphics[width=0.40\textwidth]{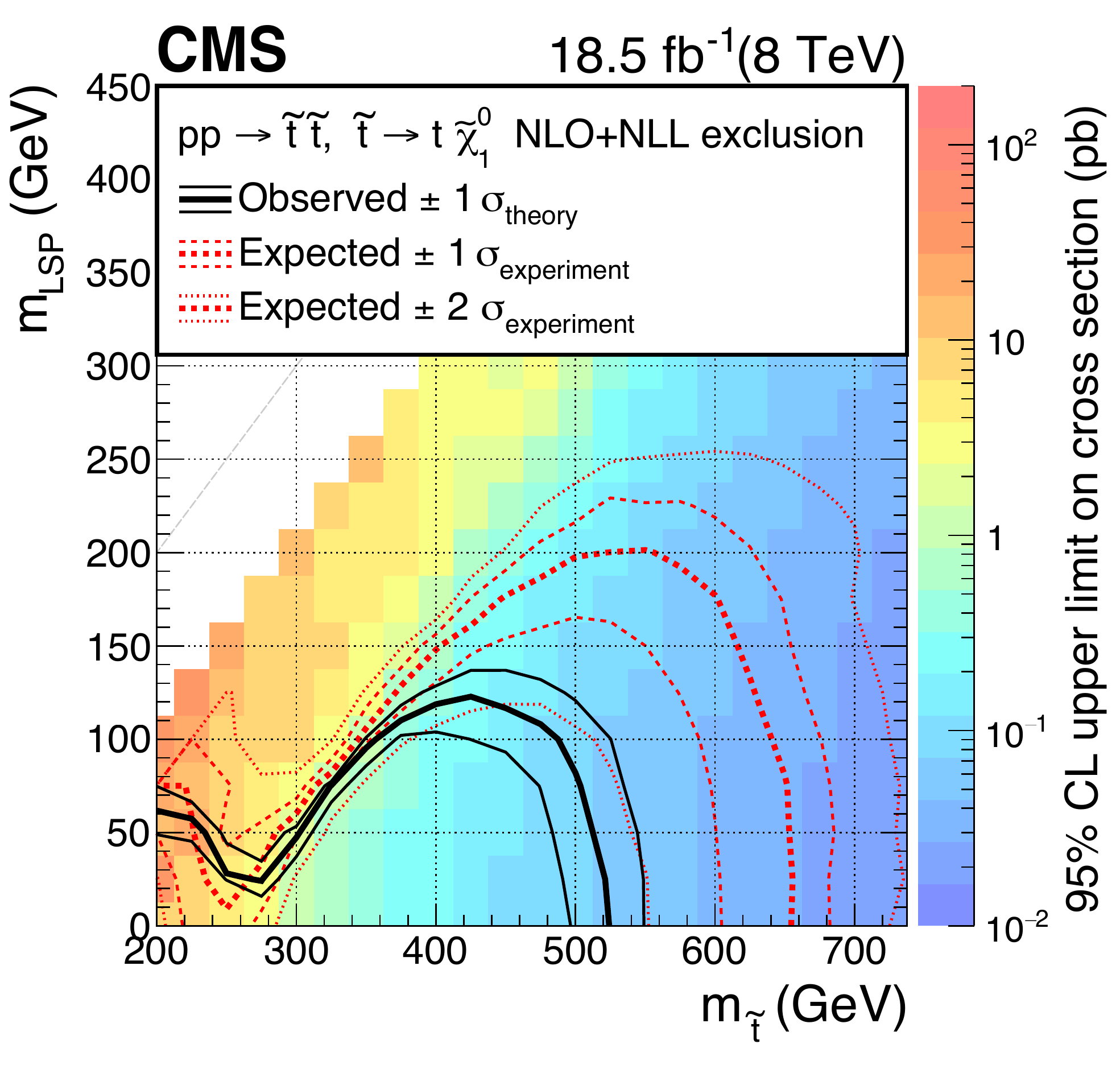}
      \includegraphics[width=0.40\textwidth]{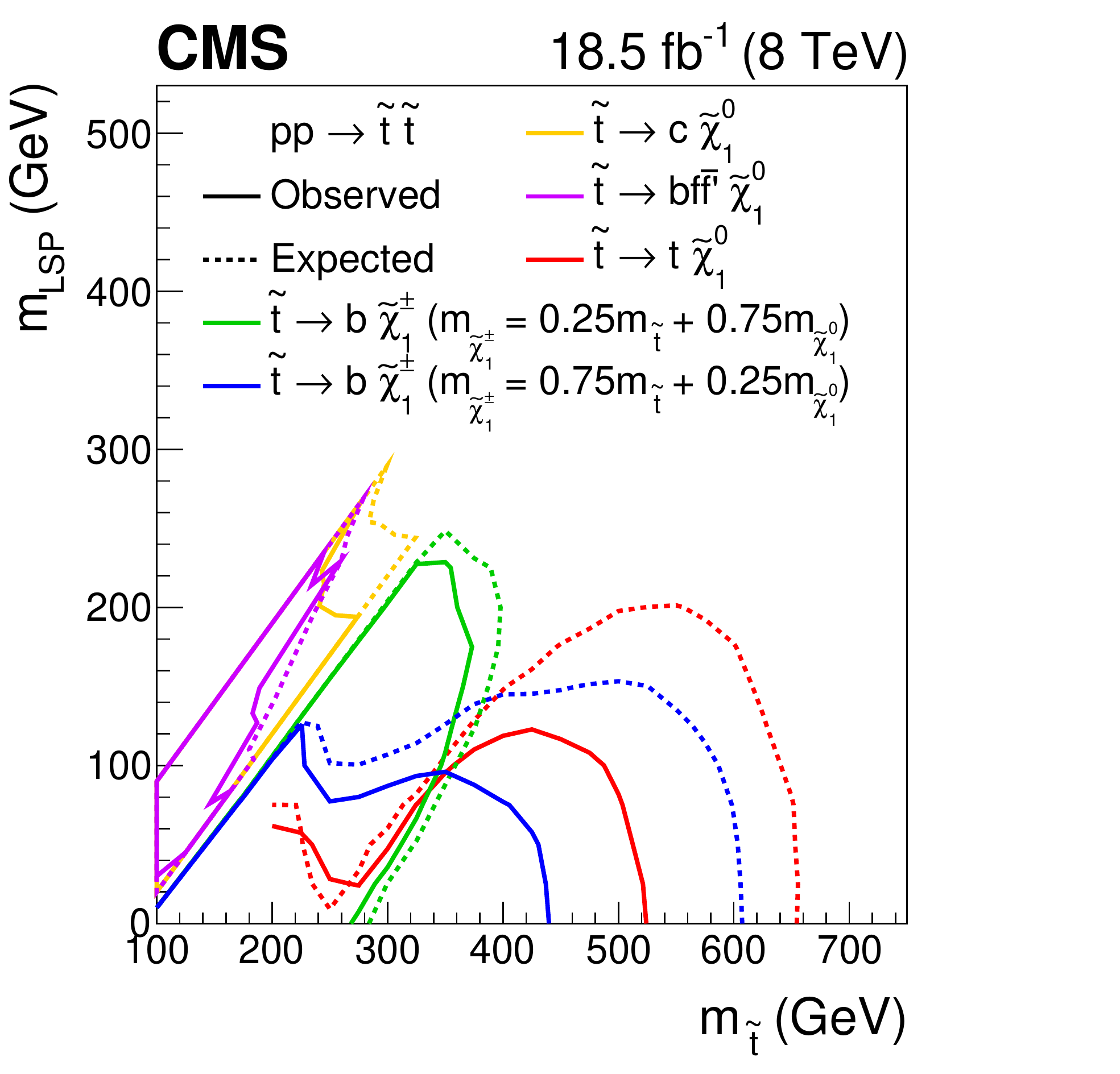}
    \caption{ Observed upper limits on the production cross section at
      95\% CL (indicated by the colour scale) as a function of the
      top squark and $\PSGczDo$ masses for
      (upper left) $\PSQt \to\PQc\PSGczDo$,
      (upper right) $\PSQt \to{\PQb \ffbp}\PSGczDo$,
      (middle left) $\PSQt \to\PQb\PSGcpmDo$ with $m_{\PSGcpmDo} = 0.25m_{\PSQt} + 0.75m_{\PSGczDo}$,
      (middle right) $\PSQt \to\PQb\PSGcpmDo$ with $m_{\PSGcpmDo} = 0.75m_{\PSQt} + 0.25m_{\PSGczDo}$, and
      (lower left) $\PSQt \to\PQt \PSGczDo$.
      The black solid thick curves indicate the observed exclusion
      assuming the NLO+NLL SUSY production cross sections; the thin
      black curves show corresponding ${\pm}1\sigma$ theoretical
      uncertainties. The red thick dashed curves indicate median
      expected exclusions and the thin dashed and dotted
      curves indicate, respectively, their ${\pm}1 \sigma$ and
      ${\pm}2\sigma$ experimental uncertainties. A summary of the
      observed (solid) and median expected (dotted) exclusion contours
      is presented (lower right). The grey dotted diagonal lines
      delimit the region for which $m_{\PSQt} > m_{\PSGczDo}$. (For
      interpretation of the references to colour in this figure
      legend, the reader is referred to the web version of this
      article.) 
      \label{fig:limits-sms}
    }
\end{figure*}

Figure~\ref{fig:limits-sms} shows the observed upper limit on the
production cross section at 95\% confidence level (CL), as a function
of the top squark and $\PSGczDo$ masses, for a range of simplified
models based on the pair production of top squarks, together with
excluded mass regions.

Figures~\ref{fig:limits-sms} (upper left and right) show the
sensitivity of this analysis to the decay modes $\PSQt \to
\PQc\PSGczDo$ and $\PSQt \to\PQb \ffbp \PSGczDo$, respectively. Models
with \dm as small as 10\GeV are considered, and the top squarks are
assumed to decay promptly. The excluded regions are determined using
the NLO+NLL cross sections for top squark pair production, assuming
that b squarks, light-flavoured squarks, and gluinos are too heavy to
be produced in the pp collisions. Also shown are the excluded regions
observed when the production cross section is changed by its
theoretical uncertainty, and the expected region of exclusion, as well
as those determined for both ${\pm}1$ and ${\pm}2$ standard deviation
($\sigma$) changes in experimental uncertainties. The range of
excluded top squark masses is sensitive to both the decay mode and
\dm. For the decay $\PSQt \to\PQc\PSGczDo$, the expected excluded
region is relatively stable as a function of \dm, with $\PSQt$ masses
below 285 and 325\GeV excluded, respectively, for $\dm = 10$ and
80\GeV. The observed exclusion, assuming the theoretical production
cross section reduced by its $1\sigma$ uncertainty, is weaker, with
$\PSQt$ masses below 240 and 260\GeV excluded for $\dm = 10$ and
80\GeV. For the decay $\PSQt \to\PQb \ffbp \PSGczDo$, the expected
excluded mass region is strongly dependent on \dm, weakening
considerably for increasing values of \dm due to the increased
momentum phase space available to leptons produced in the four-body
decay. Top squark masses below 265 and 165\GeV are excluded based on
the expected results, respectively, for $\dm = 10$ and 80\GeV. The
observed exclusion is again weaker, with masses below 225 and 130\GeV
excluded. The nonsmooth behaviour of the exclusion contours is the
result of statistical fluctuations and the sparseness of the scan over
the mass parameter space, and does not represent a kinematical effect.

Figures~\ref{fig:limits-sms} (middle left and right) show the limits
on the allowed cross section for the decay $\PSQt \to \PQb \PSGcpmDo$,
followed by a decay of the $\PSGcpmDo$ to the $\PSGczDo$ and to either
an on- or off-shell W boson, depending on the mass difference between
the $\PSGcpmDo$ and $\PSGczDo$.  For a model with $m_{\PSGcpmDo} =
0.25m_{\PSQt} + 0.75m_{\PSGczDo}$, shown in Fig.~\ref{fig:limits-sms}
(middle left), the analysis has sensitivity in the region
$m_{\PSGcpmDo} - m_{\PSGczDo} < m_{\PW}$, excluding $\PSGczDo$ masses
up to 225\GeV and $\PSQt$ masses up to 350\GeV. Models that satisfy
$m_{\PSGcpmDo} < 91.9\GeV$, or $m_{\PSGcpmDo} < 103.5\GeV$ and
$m_{\PSGcpmDo} - m_{\PSGczDo} < 5\GeV$, are already excluded by a
combination of results obtained from the ALEPH, DELPHI, L3, and OPAL
experiments at LEP~\cite{lep1, lep2}. For a model with $m_{\PSGcpmDo}
= 0.75m_{\PSQt} + 0.25m_{\PSGczDo}$, shown in
Fig.~\ref{fig:limits-sms} (middle right), $\PSQt$ masses up to $400
\GeV$ can be excluded but the reach in $\PSGczDo$ mass is reduced.

Figure~\ref{fig:limits-sms} (lower left) shows the results of the
analysis for the decay $\PSQt \to \PQt \PSGczDo$. Both two- and
three-body decays are considered, for which the latter scenario
involves an off-shell top quark. The polarizations of the top quarks
are model dependent and are non-trivial functions of the top-squark
and neutralino mixing matrices~\cite{Perelstein:2008zt}. Simulated
events of the production and decay of top squark pairs are generated
without polarization of the top quarks. Models with $m_{\PSQt} <
200\GeV$ are not considered, due to significant signal contributions
in the control regions. Top squark masses up to 500 GeV are excluded,
and $\PSGczDo$ masses up to 100 and 50\GeV are excluded for the two-
and three-body decays, respectively. As in Fig.~\ref{fig:limits-sms}
(middle right), the observed limit is around 2$\sigma$ below the
expected result for large values of $m_{\PSQt}$. This is mainly due to
an excess of observed counts in data in the $\nb=2$ categories in the
region of $500 < \scalht < 700 \GeV$, which is compatible with a
statistical fluctuation.  The observed limits lie closer to the
expected values at low top squark masses, which correspond to lower
values of \scalht for which good agreement between the data and SM
background predictions is observed.

Figure~\ref{fig:limits-sms} (lower right) presents a summary of all
the expected and observed exclusion contours and indicates that the
analysis has good sensitivity across many different decay signatures
in the $m_{\PSQt}$--$m_{\PSGczDo}$ plane. The sensitivity for these
models is typically driven by categories involving events satisfying
$\njet \geq 4$ and $1 \leq \nb \leq 2$, while events with lower \njet
and \nb multiplicities become increasingly important for nearly
mass-degenerate models.

\section{Summary}

An inclusive search for supersymmetry with the CMS detector is
reported, based on data from pp collisions collected at $\sqrt{s} =
8\TeV$, corresponding to an integrated luminosity of $18.5 \pm 0.5
\fbinv$. The final states analysed contain two or more jets with large
transverse energies and a significant imbalance in the event
transverse momentum, as expected in the production and decay of
massive squarks and gluinos. Dedicated triggers made it possible to
extend the phase space covered in this search to values of \scalht and
\HTmiss as low as 200 and 130\GeV, respectively.  These regions of low
\scalht and \HTmiss correspond to regions of phase space that are
highly populated in models with low-mass squarks and nearly degenerate
mass spectra. The signal region is binned according to \scalht, the
number of reconstructed jets, and the number of jets identified as
originating from b quarks. The sum of standard model backgrounds in
each bin is estimated from a simultaneous binned likelihood fit to the
event yields in the signal region and in \mj, \mmj, and \gj control
samples. The observed yields in the signal region are found to be in
agreement with the expected contributions from standard model
processes.

Limits are determined in the mass parameter space of simplified models
that assume the direct pair production of top squarks. A comprehensive
study of top squark decay modes is performed and interpreted in the
parameter space of the loop-induced two-body decays to the neutralino
and one c quark ($\PSQt \to \PQc\PSGczDo$); four-body decays to the
neutralino, one b quark, and an off-shell W boson ($\PSQt \to {\PQb
  \ffbp} \PSGczDo$); decays to one b quark and the lightest chargino
($\PSQt \to \PQb \PSGcpmDo$), followed by the decay of the chargino to
the lightest neutralino and an (off-shell) W boson; and the decay to a
top quark and neutralino ($\PSQt \to \PQt \PSGczDo$). In the region
$m_{\PSQt} - m_{\PSGczDo} < m_{\PW}$, top squarks with masses as large
as 260 and 225\GeV, and neutralino masses up to 240 and 215\GeV, are
excluded, respectively, for the two- and four-body decay modes. For
top squark decays to $\PQb\PSGcpmDo$, top squark masses up to 400\GeV
and neutralino masses up to 225\GeV are excluded, depending on the
mass of the chargino. For top squarks decaying to a top quark and a
neutralino, top squark masses up to 500\GeV and neutralino masses up
to 105\GeV are excluded.

In summary, the analysis provides sensitivity across a large region of
parameter space in the ($m_{\PSQt}, m_{\PSGczDo}$) plane, covering
several relevant top squark decay modes. In particular, the
application of low thresholds to maximise signal acceptance provides
sensitivity to models with compressed mass spectra. For top squark
decays to b$\PSGcpmDo$, where the W boson from the $\PSGcpmDo$ decay
is off-shell, the presented studies improve on existing limits. Mass
exclusions are reported in previously unexplored regions of the
$(m_{\PSQt}, m_{\PSGcpmDo}, m_{\PSGczDo})$ parameter space that
satisfy $100\GeV < \dm < m_\PQt$, of up to $m_{\PSQt} = 325$,
$m_{\PSGcpmDo} = 250$, and $m_{\PSGczDo} = 225\GeV$. For the region
$\dm < m_\PW$, the search provides the strongest expected mass
exclusions, up to $m_{\PSQt} = 325\GeV$, for the two-body decay $\PSQt
\to \PQc\PSGczDo$ when $30\GeV < \dm < m_\PW$.

\begin{acknowledgments}

  We congratulate our colleagues in the CERN accelerator departments
  for the excellent performance of the LHC and thank the technical and
  administrative staffs at CERN and at other CMS institutes for their
  contributions to the success of the CMS effort. In addition, we
  gratefully acknowledge the computing centres and personnel of the
  Worldwide LHC Computing Grid for delivering so effectively the
  computing infrastructure essential to our analyses. Finally, we
  acknowledge the enduring support for the construction and operation
  of the LHC and the CMS detector provided by the following funding
  agencies: BMWFW and FWF (Austria); FNRS and FWO (Belgium); CNPq,
  CAPES, FAPERJ, and FAPESP (Brazil); MES (Bulgaria); CERN; CAST Innovation Foundation, MoST,
  and NSFC (China); COLCIENCIAS (Colombia); MSES and CSF (Croatia);
  RPF (Cyprus); MoER, ERC IUT and ERDF (Estonia); Academy of Finland,
  Nokia, and HIP (Finland); CEA and CNRS/IN2P3 (France); BMBF, DFG, and
  HGF (Germany); GSRT (Greece); OTKA and NIH (Hungary); DAE and DST
  (India); IPM (Iran); SFI (Ireland); INFN (Italy); MSIP and NRF
  (Republic of Korea); LAS (Lithuania); MOE and UM (Malaysia); BUAP,
  CINVESTAV, CONACYT, Instituto de Ciencia y Tecnolog{\'i}a del Distrito Federal, SEP, and UASLP-FAI (Mexico); MBIE (New
  Zealand); PAEC (Pakistan); MSHE and NSC (Poland); FCT (Portugal);
  JINR (Dubna); MON, RosAtom, RAS and RFBR (Russia); MESTD (Serbia);
  SEIDI and CPAN (Spain); Swiss Funding Agencies (Switzerland); MST
  (Taipei); ThEPCenter, IPST, STAR and NSTDA (Thailand); TUBITAK and
  TAEK (Turkey); NASU and SFFR (Ukraine); STFC (United Kingdom); DOE
  and NSF (USA).

  Individuals have received support from the Marie-Curie programme and
  the European Research Council and EPLANET (European Union); the
  Leventis Foundation; the Alfred P. Sloan Foundation; the Alexander von
  Humboldt Foundation; the Belgian Federal Science Policy Office; the
  Fonds pour la Formation \`a la Recherche dans l'Industrie et dans
  l'Agriculture (FRIA-Belgium); the Agentschap voor Innovatie door
  Wetenschap en Technologie (IWT-Belgium); the Ministry of Education,
  Youth and Sports (MEYS) of the Czech Republic; the Council of
  Scientific and Industrial Research, India; the HOMING PLUS programme of
  the Foundation for Polish Science, cofinanced from European Union,
  Regional Development Fund; the Mobility Plus programme of the
  Ministry of Science and Higher Education (Poland); the OPUS
  programme of the National Science Centre of Poland (Poland); the Thalis and
  Aristeia programmes cofinanced by EU-ESF and the Greek NSRF; the
  National Priorities Research Program by Qatar National Research
  Fund; the Programa Clar\'in-COFUND del Principado de Asturias; the
  Rachadapisek Sompot Fund for Postdoctoral Fellowship, Chulalongkorn
  University (Thailand); the Chulalongkorn Academic into Its 2nd
  Century Project Advancement Project (Thailand); and the Welch
  Foundation, contract C-1845.

\end{acknowledgments}

\bibliography{auto_generated}

\providecommand{\href}[2]{#2}\begingroup\raggedright\begin{thebibliography}{100}%
\makeatletter
\providecommand{\hrefCMSnoop }[0]{\@secondoftwo}%
\makeatother
\providecommand{\doi}{\texttt{doi:}\begingroup \urlstyle{tt}\Url}

\bibitem{ref:SUSY-1}
\href {http://www.jetpletters.ac.ru/ps/1584/article_24309.shtml}{Y.~A. Gol'fand
  and E.~P. Likhtman, ``Extension of the Algebra of {Poincar\'e} Group
  Generators and Violation of p Invariance'',} \textit{ JETP Lett.} \textbf{
  13} (1971) 323.

\bibitem{ref:SUSY0}
\hrefCMSnoop {}{J.~Wess and B.~Zumino, ``Supergauge transformations in four
  dimensions'',} \textit{ Nucl. Phys. B} \textbf{ 70} (1974) 39,
  \href{http://dx.doi.org/10.1016/0550-3213(74)90355-1}{\doi{10.1016/0550-3213(74)90355-1}}.

\bibitem{ref:SUSY1}
\hrefCMSnoop {}{H.~P. Nilles, ``Supersymmetry, supergravity and particle
  physics'',} \textit{ Phys. Reports} \textbf{ 110} (1984) 1,
  \href{http://dx.doi.org/10.1016/0370-1573(84)90008-5}{\doi{10.1016/0370-1573(84)90008-5}}.

\bibitem{ref:SUSY2}
\hrefCMSnoop {}{H.~E. Haber and G.~L. Kane, ``The search for supersymmetry:
  Probing physics beyond the standard model'',} \textit{ Phys. Reports}
  \textbf{ 117} (1987) 75,
  \href{http://dx.doi.org/10.1016/0370-1573(85)90051-1}{\doi{10.1016/0370-1573(85)90051-1}}.

\bibitem{ref:SUSY3}
\hrefCMSnoop {}{R.~Barbieri, S.~Ferrara, and C.~A. Savoy, ``Gauge models with
  spontaneously broken local supersymmetry'',} \textit{ Phys. Lett. B} \textbf{
  119} (1982) 343,
  \href{http://dx.doi.org/10.1016/0370-2693(82)90685-2}{\doi{10.1016/0370-2693(82)90685-2}}.

\bibitem{ref:SUSY4}
\hrefCMSnoop {}{S.~Dawson, E.~Eichten, and C.~Quigg, ``Search for
  supersymmetric particles in hadron-hadron collisions'',} \textit{ Phys. Rev.
  D} \textbf{ 31} (1985) 1581,
  \href{http://dx.doi.org/10.1103/PhysRevD.31.1581}{\doi{10.1103/PhysRevD.31.1581}}.

\bibitem{ref:hierarchy1}
\hrefCMSnoop {}{E.~Witten, ``Dynamical breaking of supersymmetry'',} \textit{
  Nucl. Phys. B} \textbf{ 188} (1981) 513,
  \href{http://dx.doi.org/10.1016/0550-3213(81)90006-7}{\doi{10.1016/0550-3213(81)90006-7}}.

\bibitem{ref:hierarchy2}
\hrefCMSnoop {}{S.~Dimopoulos and H.~Georgi, ``Softly broken supersymmetry and
  {SU(5)}'',} \textit{ Nucl. Phys. B} \textbf{ 193} (1981) 150,
  \href{http://dx.doi.org/10.1016/0550-3213(81)90522-8}{\doi{10.1016/0550-3213(81)90522-8}}.

\bibitem{ref:atlashiggsdiscovery}
\hrefCMSnoop {}{{ATLAS Collaboration}, ``{Observation of a new particle in the
  search for the Standard Model Higgs boson with the ATLAS detector at the
  LHC}'',} \textit{ Phys. Lett. B} \textbf{ 716} (2012) 1,
  \href{http://dx.doi.org/10.1016/j.physletb.2012.08.020}{\doi{10.1016/j.physletb.2012.08.020}},
\href{http://www.arXiv.org/abs/1207.7214}{\texttt{arXiv:1207.7214}}.

\bibitem{ref:cmshiggsdiscoverylong}
\hrefCMSnoop {}{{CMS Collaboration}, ``{Observation of a new boson with mass
  near 125 GeV in pp collisions at $\sqrt{s} = 7$~and~8\TeV}'',} \textit{ JHEP}
  \textbf{ 06} (2013) 081,
  \href{http://dx.doi.org/10.1007/JHEP06(2013)081}{\doi{10.1007/JHEP06(2013)081}},
\href{http://www.arXiv.org/abs/1303.4571}{\texttt{arXiv:1303.4571}}.

\bibitem{ref:barbierinsusy}
\hrefCMSnoop {}{R.~Barbieri and D.~Pappadopulo, ``{S-particles at their
  naturalness limits}'',} \textit{ JHEP} \textbf{ 10} (2009) 061,
  \href{http://dx.doi.org/10.1088/1126-6708/2009/10/061}{\doi{10.1088/1126-6708/2009/10/061}},
\href{http://www.arXiv.org/abs/0906.4546}{\texttt{arXiv:0906.4546}}.

\bibitem{Farrar:1978xj}
\hrefCMSnoop {}{G.~R. Farrar and P.~Fayet, ``Phenomenology of the production,
  decay, and detection of new hadronic states associated with supersymmetry'',}
  \textit{ Phys. Lett. B} \textbf{ 76} (1978) 575,
\href{http://dx.doi.org/10.1016/0370-2693(78)90858-4}{\doi{10.1016/0370-2693(78)90858-4}}.

\bibitem{Boehm:1999tr}
\hrefCMSnoop {}{C.~Boehm, A.~Djouadi, and Y.~Mambrini, ``{Decays of the
  lightest top squark}'',} \textit{ Phys. Rev. D} \textbf{ 61} (2000) 095006,
  \href{http://dx.doi.org/10.1103/PhysRevD.61.095006}{\doi{10.1103/PhysRevD.61.095006}},
\href{http://www.arXiv.org/abs/hep-ph/9907428}{\texttt{arXiv:hep-ph/9907428}}.

\bibitem{Boehm:1999bj}
\hrefCMSnoop {}{C.~Boehm, A.~Djouadi, and M.~Drees, ``{Light scalar top quarks
  and supersymmetric dark matter}'',} \textit{ Phys. Rev. D} \textbf{ 62}
  (2000) 035012,
  \href{http://dx.doi.org/10.1103/PhysRevD.62.035012}{\doi{10.1103/PhysRevD.62.035012}},
\href{http://www.arXiv.org/abs/hep-ph/9911496}{\texttt{arXiv:hep-ph/9911496}}.

\bibitem{Balazs:2004bu}
\hrefCMSnoop {}{C.~Balazs, M.~S. Carena, and C.~E.~M. Wagner, ``Dark matter,
  light stops and electroweak baryogenesis'',} \textit{ Phys. Rev. D} \textbf{
  70} (2004) 015007,
  \href{http://dx.doi.org/10.1103/PhysRevD.70.015007}{\doi{10.1103/PhysRevD.70.015007}},
\href{http://www.arXiv.org/abs/hep-ph/0403224}{\texttt{arXiv:hep-ph/0403224}}.

\bibitem{Martin:2007gf}
\hrefCMSnoop {}{S.~P. Martin, ``{Compressed supersymmetry and natural
  neutralino dark matter from top squark-mediated annihilation to top
  quarks}'',} \textit{ Phys. Rev. D} \textbf{ 75} (2007) 115005,
  \href{http://dx.doi.org/10.1103/PhysRevD.75.115005}{\doi{10.1103/PhysRevD.75.115005}},
\href{http://www.arXiv.org/abs/hep-ph/0703097}{\texttt{arXiv:hep-ph/0703097}}.

\bibitem{Martin:2007hn}
\hrefCMSnoop {}{S.~P. Martin, ``{Top squark-mediated annihilation scenario and
  direct detection of dark matter in compressed supersymmetry}'',} \textit{
  Phys. Rev. D} \textbf{ 76} (2007) 095005,
  \href{http://dx.doi.org/10.1103/PhysRevD.76.095005}{\doi{10.1103/PhysRevD.76.095005}},
\href{http://www.arXiv.org/abs/0707.2812}{\texttt{arXiv:0707.2812}}.

\bibitem{Carena:2008mj}
\hrefCMSnoop {}{M.~Carena, A.~Freitas, and C.~E.~M. Wagner, ``Light stop
  searches at the {LHC} in events with one hard photon or jet and missing
  energy'',} \textit{ JHEP} \textbf{ 10} (2008) 109,
  \href{http://dx.doi.org/10.1088/1126-6708/2008/10/109}{\doi{10.1088/1126-6708/2008/10/109}},
\href{http://www.arXiv.org/abs/0808.2298}{\texttt{arXiv:0808.2298}}.

\bibitem{Grober:2014aha}
\hrefCMSnoop {}{R.~Grober, M.~M. Muhlleitner, E.~Popenda, and A.~Wlotzka,
  ``Light stop decays: implications for {LHC} searches'',} \textit{ Eur. Phys.
  J. C} \textbf{ 75} (2015) 420,
  \href{http://dx.doi.org/10.1140/epjc/s10052-015-3626-z}{\doi{10.1140/epjc/s10052-015-3626-z}},
\href{http://www.arXiv.org/abs/1408.4662}{\texttt{arXiv:1408.4662}}.

\bibitem{Grober:2015fia}
\hrefCMSnoop {}{R.~Grober, M.~Muhlleitner, E.~Popenda, and A.~Wlotzka, ``{Light
  stop decays into $\text{Wb}\chiz_1$ near the kinematic threshold}'',}
  \textit{ Phys. Lett. B} \textbf{ 747} (2015) 144,
  \href{http://dx.doi.org/10.1016/j.physletb.2015.05.060}{\doi{10.1016/j.physletb.2015.05.060}},
\href{http://www.arXiv.org/abs/1502.05935}{\texttt{arXiv:1502.05935}}.

\bibitem{atlas-13}
\hrefCMSnoop {}{{ATLAS Collaboration}, ``{Search for new phenomena in final
  states with an energetic jet and large missing transverse momentum in $pp$
  collisions at $\sqrt{s}=13\TeV$ using the ATLAS detector}'',} \textit{ Phys.
  Rev. D} \textbf{ 94} (2016) 032005,
  \href{http://dx.doi.org/10.1103/PhysRevD.94.032005}{\doi{10.1103/PhysRevD.94.032005}},
\href{http://www.arXiv.org/abs/1604.07773}{\texttt{arXiv:1604.07773}}.

\bibitem{atlas-6}
\hrefCMSnoop {}{{ATLAS Collaboration}, ``{Search for pair-produced
  third-generation squarks decaying via charm quarks or in compressed
  supersymmetric scenarios in pp collisions at $\sqrt{s} = 8\TeV$ with the
  ATLAS detector}'',} \textit{ Phys. Rev. D} \textbf{ 90} (2014) 052008,
  \href{http://dx.doi.org/10.1103/PhysRevD.90.052008}{\doi{10.1103/PhysRevD.90.052008}},
\href{http://www.arXiv.org/abs/1407.0608}{\texttt{arXiv:1407.0608}}.

\bibitem{cms-9}
\hrefCMSnoop {}{{CMS Collaboration}, ``{Searches for third-generation squark
  production in fully hadronic final states in proton-proton collisions at $
  \sqrt{s} = 8\TeV$}'',} \textit{ JHEP} \textbf{ 06} (2015) 116,
  \href{http://dx.doi.org/10.1007/JHEP06(2015)116}{\doi{10.1007/JHEP06(2015)116}},
\href{http://www.arXiv.org/abs/1503.08037}{\texttt{arXiv:1503.08037}}.

\bibitem{lumi}
\href {http://cdsweb.cern.ch/record/1482193}{{CMS Collaboration}, ``CMS
  Luminosity Based on Pixel Cluster Counting - Summer 2012 Update'',} CMS
  Physics Analysis Summary CMS-PAS-LUM-12-001, 2012.

\bibitem{RA1Paper}
\hrefCMSnoop {}{{CMS Collaboration}, ``Search for supersymmetry in pp
  collisions at 7 {TeV} in events with jets and missing transverse energy'',}
  \textit{ Phys. Lett. B} \textbf{ 698} (2011) 196,
  \href{http://dx.doi.org/10.1016/j.physletb.2011.03.021}{\doi{10.1016/j.physletb.2011.03.021}},
\href{http://www.arXiv.org/abs/1101.1628}{\texttt{arXiv:1101.1628}}.

\bibitem{RA1Paper2011}
\hrefCMSnoop {}{{CMS Collaboration}, ``Search for Supersymmetry at the {LHC} in
  Events with Jets and Missing Transverse Energy'',} \textit{ Phys. Rev. Lett.}
  \textbf{ 107} (2011) 221804,
  \href{http://dx.doi.org/10.1103/PhysRevLett.107.221804}{\doi{10.1103/PhysRevLett.107.221804}},
\href{http://www.arXiv.org/abs/1109.2352}{\texttt{arXiv:1109.2352}}.

\bibitem{RA1Paper2011FULL}
\hrefCMSnoop {}{{CMS Collaboration}, ``{Search for supersymmetry in final
  states with missing transverse energy and 0, 1, 2, or at least 3 b-quark jets
  in 7 TeV pp collisions using the variable $\alpha_\text{T}$}'',} \textit{
  JHEP} \textbf{ 01} (2013) 077,
  \href{http://dx.doi.org/10.1007/JHEP01(2013)077}{\doi{10.1007/JHEP01(2013)077}},
\href{http://www.arXiv.org/abs/1210.8115}{\texttt{arXiv:1210.8115}}.

\bibitem{RA1Paper2012}
\hrefCMSnoop {}{{CMS Collaboration}, ``{Search for supersymmetry in hadronic
  final states with missing transverse energy using the variables
  $\alpha_\text{T}$ and b-quark multiplicity in pp collisions at $\sqrt{s} =
  8\TeV$}'',} \textit{ Eur. Phys. J. C} \textbf{ 73} (2013) 2568,
  \href{http://dx.doi.org/10.1140/epjc/s10052-013-2568-6}{\doi{10.1140/epjc/s10052-013-2568-6}},
\href{http://www.arXiv.org/abs/1303.2985}{\texttt{arXiv:1303.2985}}.

\bibitem{atlas-0}
\hrefCMSnoop {}{{ATLAS Collaboration}, ``{Multi-channel search for squarks and
  gluinos in $\sqrt{s} = 7\TeV$ pp collisions with the ATLAS detector}'',}
  \textit{ Eur. Phys. J. C} \textbf{ 73} (2013) 2362,
  \href{http://dx.doi.org/10.1140/epjc/s10052-013-2362-5}{\doi{10.1140/epjc/s10052-013-2362-5}},
\href{http://www.arXiv.org/abs/1212.6149}{\texttt{arXiv:1212.6149}}.

\bibitem{atlas-1}
\hrefCMSnoop {}{{ATLAS Collaboration}, ``Search for a Supersymmetric Partner to
  the Top Quark in Final States with Jets and Missing Transverse Momentum at
  {$\sqrt{s} = 7\TeV$} with the {ATLAS} Detector'',} \textit{ Phys. Rev. Lett.}
  \textbf{ 109} (2012) 211802,
  \href{http://dx.doi.org/10.1103/PhysRevLett.109.211802}{\doi{10.1103/PhysRevLett.109.211802}},
\href{http://www.arXiv.org/abs/1208.1447}{\texttt{arXiv:1208.1447}}.

\bibitem{atlas-2}
\hrefCMSnoop {}{{ATLAS Collaboration}, ``{Search for squarks and gluinos using
  final states with jets and missing transverse momentum with the ATLAS
  detector in $\sqrt{s} = 7\TeV$ proton-proton collisions}'',} \textit{ Phys.
  Lett. B} \textbf{ 710} (2012) 67,
  \href{http://dx.doi.org/10.1016/j.physletb.2012.02.051}{\doi{10.1016/j.physletb.2012.02.051}},
\href{http://www.arXiv.org/abs/1109.6572}{\texttt{arXiv:1109.6572}}.

\bibitem{atlas-3}
\hrefCMSnoop {}{{ATLAS Collaboration}, ``{Search for top and bottom squarks
  from gluino pair production in final states with missing transverse energy
  and at least three b-jets with the ATLAS detector}'',} \textit{ Eur. Phys. J.
  C} \textbf{ 72} (2012) 2174,
  \href{http://dx.doi.org/10.1140/epjc/s10052-012-2174-z}{\doi{10.1140/epjc/s10052-012-2174-z}},
\href{http://www.arXiv.org/abs/1207.4686}{\texttt{arXiv:1207.4686}}.

\bibitem{atlas-4}
\hrefCMSnoop {}{{ATLAS Collaboration}, ``{Hunt for new phenomena using large
  jet multiplicities and missing transverse momentum with ATLAS in 4.7
  fb$^{-1}$ of $\sqrt{s} = 7\TeV$ proton-proton collisions}'',} \textit{ JHEP}
  \textbf{ 07} (2012) 167,
  \href{http://dx.doi.org/10.1007/JHEP07(2012)167}{\doi{10.1007/JHEP07(2012)167}},
\href{http://www.arXiv.org/abs/1206.1760}{\texttt{arXiv:1206.1760}}.

\bibitem{atlas-5}
\hrefCMSnoop {}{{ATLAS Collaboration}, ``Search for Scalar Bottom Quark Pair
  Production with the {ATLAS} Detector in pp Collisions at {$\sqrt{s} =
  7\TeV$}'',} \textit{ Phys. Rev. Lett.} \textbf{ 108} (2012) 181802,
  \href{http://dx.doi.org/10.1103/PhysRevLett.108.181802}{\doi{10.1103/PhysRevLett.108.181802}},
\href{http://www.arXiv.org/abs/1112.3832}{\texttt{arXiv:1112.3832}}.

\bibitem{atlas-11}
\hrefCMSnoop {}{{ATLAS Collaboration}, ``{Search for new phenomena in final
  states with large jet multiplicities and missing transverse momentum using
  $\sqrt{s} = 7\TeV$ pp collisions with the ATLAS detector}'',} \textit{ JHEP}
  \textbf{ 11} (2011) 099,
  \href{http://dx.doi.org/10.1007/JHEP11(2011)099}{\doi{10.1007/JHEP11(2011)099}},
\href{http://www.arXiv.org/abs/1110.2299}{\texttt{arXiv:1110.2299}}.

\bibitem{atlas-7}
\hrefCMSnoop {}{{ATLAS Collaboration}, ``{Search for strong production of
  supersymmetric particles in final states with missing transverse momentum and
  at least three b-jets at $\sqrt{s} = 8\TeV$ proton-proton collisions with the
  ATLAS detector}'',} \textit{ JHEP} \textbf{ 10} (2014) 24,
  \href{http://dx.doi.org/10.1007/JHEP10(2014)024}{\doi{10.1007/JHEP10(2014)024}},
\href{http://www.arXiv.org/abs/1407.0600}{\texttt{arXiv:1407.0600}}.

\bibitem{atlas-8}
\hrefCMSnoop {}{{ATLAS Collaboration}, ``{Search for squarks and gluinos with
  the ATLAS detector in final states with jets and missing transverse momentum
  using $\sqrt{s} = 8\TeV$ proton--proton collision data}'',} \textit{ JHEP}
  \textbf{ 09} (2014) 176,
  \href{http://dx.doi.org/10.1007/JHEP09(2014)176}{\doi{10.1007/JHEP09(2014)176}},
\href{http://www.arXiv.org/abs/1405.7875}{\texttt{arXiv:1405.7875}}.

\bibitem{atlas-9}
\hrefCMSnoop {}{{ATLAS Collaboration}, ``{Search for direct third-generation
  squark pair production in final states with missing transverse momentum and
  two b-jets in $\sqrt{s} = 8\TeV$ pp collisions with the ATLAS detector}'',}
  \textit{ JHEP} \textbf{ 10} (2013) 189,
  \href{http://dx.doi.org/10.1007/JHEP10(2013)189}{\doi{10.1007/JHEP10(2013)189}},
\href{http://www.arXiv.org/abs/1308.2631}{\texttt{arXiv:1308.2631}}.

\bibitem{atlas-10}
\hrefCMSnoop {}{{ATLAS Collaboration}, ``{Search for direct pair production of
  the top squark in all-hadronic final states in proton-proton collisions at
  $\sqrt{s} = 8\TeV$ with the ATLAS detector}'',} \textit{ JHEP} \textbf{ 09}
  (2014) 015,
  \href{http://dx.doi.org/10.1007/JHEP09(2014)015}{\doi{10.1007/JHEP09(2014)015}},
\href{http://www.arXiv.org/abs/1406.1122}{\texttt{arXiv:1406.1122}}.

\bibitem{atlas-12}
\hrefCMSnoop {}{{ATLAS Collaboration}, ``{Search for new phenomena in final
  states with large jet multiplicities and missing transverse momentum with
  ATLAS using $\sqrt{s} = 13\TeV$ proton--proton collisions}'',} \textit{ Phys.
  Lett. B} \textbf{ 757} (2016) 334,
  \href{http://dx.doi.org/10.1016/j.physletb.2016.04.005}{\doi{10.1016/j.physletb.2016.04.005}},
\href{http://www.arXiv.org/abs/1602.06194}{\texttt{arXiv:1602.06194}}.

\bibitem{Aad:2016eki}
\hrefCMSnoop {}{{ATLAS Collaboration}, ``{Search for pair production of gluinos
  decaying via stop and sbottom in events with $b$-jets and large missing
  transverse momentum in $pp$ collisions at $\sqrt{s} = 13\TeV$ with the ATLAS
  detector}'',} \textit{ Phys. Rev. D} \textbf{ 94} (2016) 032003,
  \href{http://dx.doi.org/10.1103/PhysRevD.94.032003}{\doi{10.1103/PhysRevD.94.032003}},
\href{http://www.arXiv.org/abs/1605.09318}{\texttt{arXiv:1605.09318}}.

\bibitem{Aaboud:2016zdn}
\hrefCMSnoop {}{{ATLAS Collaboration}, ``{Search for squarks and gluinos in
  final states with jets and missing transverse momentum at $\sqrt{s} = 13\TeV$
  with the ATLAS detector}'',} \textit{ Eur. Phys. J. C} \textbf{ 76} (2016)
  392,
  \href{http://dx.doi.org/10.1140/epjc/s10052-016-4184-8}{\doi{10.1140/epjc/s10052-016-4184-8}},
\href{http://www.arXiv.org/abs/1605.03814}{\texttt{arXiv:1605.03814}}.

\bibitem{cms-1}
\hrefCMSnoop {}{{CMS Collaboration}, ``{Search for supersymmetry in events with
  b-quark jets and missing transverse energy in pp collisions at 7 TeV}'',}
  \textit{ Phys. Rev. D} \textbf{ 86} (2012) 072010,
  \href{http://dx.doi.org/10.1103/PhysRevD.86.072010}{\doi{10.1103/PhysRevD.86.072010}},
\href{http://www.arXiv.org/abs/1208.4859}{\texttt{arXiv:1208.4859}}.

\bibitem{cms-2}
\hrefCMSnoop {}{{CMS Collaboration}, ``{Search for supersymmetry in hadronic
  final states using M$_{T2}$ in pp collisions at $\sqrt{s} = 7\TeV$}'',}
  \textit{ JHEP} \textbf{ 10} (2012) 018,
  \href{http://dx.doi.org/10.1007/JHEP10(2012)018}{\doi{10.1007/JHEP10(2012)018}},
\href{http://www.arXiv.org/abs/1207.1798}{\texttt{arXiv:1207.1798}}.

\bibitem{cms-3}
\hrefCMSnoop {}{{CMS Collaboration}, ``Search for New Physics in the Multijet
  and Missing Transverse Momentum Final State in Proton-Proton Collisions at
  {$\sqrt{s} = 7\TeV$}'',} \textit{ Phys. Rev. Lett.} \textbf{ 109} (2012)
  17180,
  \href{http://dx.doi.org/10.1103/PhysRevLett.109.171803}{\doi{10.1103/PhysRevLett.109.171803}},
\href{http://www.arXiv.org/abs/1207.1898}{\texttt{arXiv:1207.1898}}.

\bibitem{cms-4}
\hrefCMSnoop {}{{CMS Collaboration}, ``{Inclusive search for squarks and
  gluinos in pp collisions at $\sqrt{s} = 7\TeV$}'',} \textit{ Phys. Rev. D}
  \textbf{ 85} (2012) 012004,
  \href{http://dx.doi.org/10.1103/PhysRevD.85.012004}{\doi{10.1103/PhysRevD.85.012004}},
\href{http://www.arXiv.org/abs/1107.1279}{\texttt{arXiv:1107.1279}}.

\bibitem{cms-8}
\hrefCMSnoop {}{{CMS Collaboration}, ``{Search for supersymmetry with razor
  variables in pp collisions at $\sqrt{s} = 7\TeV$}'',} \textit{ Phys. Rev. D}
  \textbf{ 90} (2014) 112001,
  \href{http://dx.doi.org/10.1103/PhysRevD.90.112001}{\doi{10.1103/PhysRevD.90.112001}},
\href{http://www.arXiv.org/abs/1405.3961}{\texttt{arXiv:1405.3961}}.

\bibitem{cms-11}
\hrefCMSnoop {}{{CMS Collaboration}, ``{Inclusive search for supersymmetry
  using the razor variables in pp collisions at $\sqrt{s} = 7\TeV$}'',}
  \textit{ Phys. Rev. Lett.} \textbf{ 111} (2013) 081802,
  \href{http://dx.doi.org/10.1103/PhysRevLett.111.081802}{\doi{10.1103/PhysRevLett.111.081802}},
\href{http://www.arXiv.org/abs/1212.6961}{\texttt{arXiv:1212.6961}}.

\bibitem{cms-5}
\hrefCMSnoop {}{{CMS Collaboration}, ``Search for supersymmetry using razor
  variables in events with $b$-tagged jets in $pp$ collisions at {$\sqrt{s} =
  8\TeV$}'',} \textit{ Phys. Rev. D} \textbf{ 91} (2015) 052018,
  \href{http://dx.doi.org/10.1103/PhysRevD.91.052018}{\doi{10.1103/PhysRevD.91.052018}},
\href{http://www.arXiv.org/abs/1502.00300}{\texttt{arXiv:1502.00300}}.

\bibitem{cms-6}
\hrefCMSnoop {}{{CMS Collaboration}, ``{Search for new physics in the multijet
  and missing transverse momentum final state in proton-proton collisions at
  $\sqrt{s} = 8\TeV$}'',} \textit{ JHEP} \textbf{ 06} (2014) 055,
  \href{http://dx.doi.org/10.1007/JHEP06(2014)055}{\doi{10.1007/JHEP06(2014)055}},
\href{http://www.arXiv.org/abs/1402.4770}{\texttt{arXiv:1402.4770}}.

\bibitem{cms-7}
\hrefCMSnoop {}{{CMS Collaboration}, ``{Search for gluino mediated bottom- and
  top-squark production in multijet final states in pp collisions at 8 TeV}'',}
  \textit{ Phys. Lett. B} \textbf{ 725} (2013) 243,
  \href{http://dx.doi.org/10.1016/j.physletb.2013.06.058}{\doi{10.1016/j.physletb.2013.06.058}},
\href{http://www.arXiv.org/abs/1305.2390}{\texttt{arXiv:1305.2390}}.

\bibitem{cms-10}
\hrefCMSnoop {}{{CMS Collaboration}, ``{Searches for supersymmetry using the
  M$_{T2}$ variable in hadronic events produced in pp collisions at 8 TeV}'',}
  \textit{ JHEP} \textbf{ 05} (2015) 078,
  \href{http://dx.doi.org/10.1007/JHEP05(2015)078}{\doi{10.1007/JHEP05(2015)078}},
\href{http://www.arXiv.org/abs/1502.04358}{\texttt{arXiv:1502.04358}}.

\bibitem{cms-12}
\hrefCMSnoop {}{{CMS Collaboration}, ``{Search for supersymmetry in the
  multijet and missing transverse momentum final state in pp collisions at 13
  TeV}'',} \textit{ Phys. Lett. B} \textbf{ 758} (2016) 152,
  \href{http://dx.doi.org/10.1016/j.physletb.2016.05.002}{\doi{10.1016/j.physletb.2016.05.002}},
\href{http://www.arXiv.org/abs/1602.06581}{\texttt{arXiv:1602.06581}}.

\bibitem{cms-13}
\hrefCMSnoop {}{{CMS Collaboration}, ``{Search for new physics with the
  M$_{T2}$ variable in all-jets final states produced in pp collisions at
  $\sqrt{s} = 13\TeV$}'',} \textit{ JHEP} \textbf{ 10} (2016) 006,
  \href{http://dx.doi.org/10.1007/JHEP10(2016)006}{\doi{10.1007/JHEP10(2016)006}},
\href{http://www.arXiv.org/abs/1603.04053}{\texttt{arXiv:1603.04053}}.

\bibitem{Aad:2015pfx}
\hrefCMSnoop {}{{ATLAS Collaboration}, ``{ATLAS Run 1 searches for direct pair
  production of third-generation squarks at the Large Hadron Collider}'',}
  \textit{ Eur. Phys. J. C} \textbf{ 75} (2015) 510,
  \href{http://dx.doi.org/10.1140/epjc/s10052-015-3726-9}{\doi{10.1140/epjc/s10052-015-3726-9}},
  \href{http://www.arXiv.org/abs/1506.08616}{\texttt{arXiv:1506.08616}}.
[Erratum: \DOI{10.1140/epjc/s10052-016-3935-x}].

\bibitem{Khachatryan:2016pup}
\hrefCMSnoop {}{{CMS Collaboration}, ``{Search for direct pair production of
  scalar top quarks in the single- and dilepton channels in proton-proton
  collisions at $\sqrt{s} = 8\TeV$}'',} \textit{ JHEP} \textbf{ 07} (2016) 027,
  \href{http://dx.doi.org/10.1007/JHEP07(2016)027}{\doi{10.1007/JHEP07(2016)027}},
\href{http://www.arXiv.org/abs/1602.03169}{\texttt{arXiv:1602.03169}}.

\bibitem{Aaboud:2016lwz}
\hrefCMSnoop {}{{ATLAS Collaboration}, ``{Search for top squarks in final
  states with one isolated lepton, jets, and missing transverse momentum in
  $\sqrt{s} = 13\TeV$ pp collisions with the ATLAS detector}'',} \textit{ Phys.
  Rev. D} \textbf{ 94} (2016) 052009,
  \href{http://dx.doi.org/10.1103/PhysRevD.94.052009}{\doi{10.1103/PhysRevD.94.052009}},
\href{http://www.arXiv.org/abs/1606.03903}{\texttt{arXiv:1606.03903}}.

\bibitem{Alwall:2008ag}
\hrefCMSnoop {}{J.~Alwall, P.~Schuster, and N.~Toro, ``Simplified models for a
  first characterization of new physics at the {LHC}'',} \textit{ Phys. Rev. D}
  \textbf{ 79} (2009) 075020,
  \href{http://dx.doi.org/10.1103/PhysRevD.79.075020}{\doi{10.1103/PhysRevD.79.075020}},
\href{http://www.arXiv.org/abs/0810.3921}{\texttt{arXiv:0810.3921}}.

\bibitem{Alwall:2008va}
\hrefCMSnoop {}{J.~Alwall, M.-P. Le, M.~Lisanti, and J.~G. Wacker,
  ``Model-independent jets plus missing energy searches'',} \textit{ Phys. Rev.
  D} \textbf{ 79} (2009) 015005,
  \href{http://dx.doi.org/10.1103/PhysRevD.79.015005}{\doi{10.1103/PhysRevD.79.015005}},
\href{http://www.arXiv.org/abs/0809.3264}{\texttt{arXiv:0809.3264}}.

\bibitem{sms}
\hrefCMSnoop {}{{LHC New Physics Working Group} Collaboration, ``Simplified
  models for {LHC} new physics searches'',} \textit{ J. Phys. G} \textbf{ 39}
  (2012) 105005,
  \href{http://dx.doi.org/10.1088/0954-3899/39/10/105005}{\doi{10.1088/0954-3899/39/10/105005}},
\href{http://www.arXiv.org/abs/1105.2838}{\texttt{arXiv:1105.2838}}.

\bibitem{CMS:2012ooa}
\href {https://cds.cern.ch/record/1480607}{{CMS Collaboration}, ``{Data Parking
  and Data Scouting at the CMS Experiment}'',} {CMS Detector Performance Note}
  CMS-DP-2012-022, CERN-CMS-DP-2012-022, 2012.

\bibitem{TRK-11-001}
\hrefCMSnoop {}{{CMS Collaboration}, ``{Description and performance of track
  and primary-vertex reconstruction with the CMS tracker}'',} \textit{ JINST}
  \textbf{ 9} (2014) P10009,
  \href{http://dx.doi.org/10.1088/1748-0221/9/10/P10009}{\doi{10.1088/1748-0221/9/10/P10009}},
\href{http://www.arXiv.org/abs/1405.6569}{\texttt{arXiv:1405.6569}}.

\bibitem{Chatrchyan:2012xi}
\hrefCMSnoop {}{{CMS Collaboration}, ``Performance of {CMS} muon reconstruction
  in pp collision events at {$\sqrt{s} = 7$\TeV}'',} \textit{ J. Instrum.}
  \textbf{ 7} (2012) P10002,
  \href{http://dx.doi.org/10.1088/1748-0221/7/10/P10002}{\doi{10.1088/1748-0221/7/10/P10002}}.

\bibitem{Chatrchyan:2008zzk}
\hrefCMSnoop {}{{CMS Collaboration}, ``The {CMS} experiment at the {CERN
  LHC}'',} \textit{ JINST} \textbf{ 3} (2008) S08004,
  \href{http://dx.doi.org/10.1088/1748-0221/3/08/S08004}{\doi{10.1088/1748-0221/3/08/S08004}}.

\bibitem{Randall:2008rw}
\hrefCMSnoop {}{L.~Randall and D.~Tucker-Smith, ``Dijet Searches for
  Supersymmetry at the {Large Hadron Collider}'',} \textit{ Phys. Rev. Lett.}
  \textbf{ 101} (2008) 221803,
  \href{http://dx.doi.org/10.1103/PhysRevLett.101.221803}{\doi{10.1103/PhysRevLett.101.221803}},
\href{http://www.arXiv.org/abs/0806.1049}{\texttt{arXiv:0806.1049}}.

\bibitem{Khachatryan:2015hwa}
\hrefCMSnoop {}{{CMS Collaboration}, ``Performance of electron reconstruction
  and selection with the {CMS} detector in proton-proton collisions at
  {$\sqrt{s} = 8\TeV$}'',} \textit{ JINST} \textbf{ 10} (2015) P06005,
  \href{http://dx.doi.org/10.1088/1748-0221/10/06/P06005}{\doi{10.1088/1748-0221/10/06/P06005}},
\href{http://www.arXiv.org/abs/1502.02701}{\texttt{arXiv:1502.02701}}.

\bibitem{single-lepton-stop}
\hrefCMSnoop {}{{CMS Collaboration}, ``{Search for top-squark pair production
  in the single-lepton final state in pp collisions at $\sqrt{s} = 8\TeV$}'',}
  \textit{ Eur. Phys. J. C} \textbf{ 73} (2013) 2677,
  \href{http://dx.doi.org/10.1140/epjc/s10052-013-2677-2}{\doi{10.1140/epjc/s10052-013-2677-2}},
\href{http://www.arXiv.org/abs/1308.1586}{\texttt{arXiv:1308.1586}}.

\bibitem{Khachatryan:2015iwa}
\hrefCMSnoop {}{{CMS Collaboration}, ``Performance of photon reconstruction and
  identification with the {CMS} detector in proton-proton collisions at
  {$\sqrt{s} = 8\TeV$}'',} \textit{ JINST} \textbf{ 10} (2015) P08010,
  \href{http://dx.doi.org/10.1088/1748-0221/10/08/P08010}{\doi{10.1088/1748-0221/10/08/P08010}},
\href{http://www.arXiv.org/abs/1502.02702}{\texttt{arXiv:1502.02702}}.

\bibitem{antikt}
\hrefCMSnoop {}{M.~Cacciari, G.~P. Salam, and G.~Soyez, ``{The anti-$k_t$ jet
  clustering algorithm}'',} \textit{ JHEP} \textbf{ 04} (2008) 063,
  \href{http://dx.doi.org/10.1088/1126-6708/2008/04/063}{\doi{10.1088/1126-6708/2008/04/063}},
\href{http://www.arXiv.org/abs/0802.1189}{\texttt{arXiv:0802.1189}}.

\bibitem{Chatrchyan:2011ds}
\hrefCMSnoop {}{{CMS Collaboration}, ``Determination of jet energy calibration
  and transverse momentum resolution in {CMS}'',} \textit{ JINST} \textbf{ 6}
  (2011) P11002,
  \href{http://dx.doi.org/10.1088/1748-0221/6/11/P11002}{\doi{10.1088/1748-0221/6/11/P11002}},
\href{http://www.arXiv.org/abs/1107.4277}{\texttt{arXiv:1107.4277}}.

\bibitem{Chatrchyan:2012jua}
\hrefCMSnoop {}{{CMS Collaboration}, ``{Identification of b-quark jets with the
  CMS experiment}'',} \textit{ JINST} \textbf{ 8} (2013) P04013,
  \href{http://dx.doi.org/10.1088/1748-0221/8/04/P04013}{\doi{10.1088/1748-0221/8/04/P04013}},
\href{http://www.arXiv.org/abs/1211.4462}{\texttt{arXiv:1211.4462}}.

\bibitem{Chatrchyan:2009hy}
\hrefCMSnoop {}{{CMS Collaboration}, ``identification and filtering of
  uncharacteristic noise in the {CMS} hadron calorimeter'',} \textit{ JINST}
  \textbf{ 5} (2010) T03014,
  \href{http://dx.doi.org/10.1088/1748-0221/5/03/T03014}{\doi{10.1088/1748-0221/5/03/T03014}},
\href{http://www.arXiv.org/abs/0911.4881}{\texttt{arXiv:0911.4881}}.

\bibitem{cms-met}
\hrefCMSnoop {}{{CMS Collaboration}, ``{Missing transverse energy performance
  of the CMS detector}'',} \textit{ JINST} \textbf{ 6} (2011) P09001,
  \href{http://dx.doi.org/10.1088/1748-0221/6/09/P09001}{\doi{10.1088/1748-0221/6/09/P09001}},
\href{http://www.arXiv.org/abs/1106.5048}{\texttt{arXiv:1106.5048}}.

\bibitem{CMS-PAS-PFT-09-001}
\href {http://cdsweb.cern.ch/record/1194487}{{CMS Collaboration},
  ``Particle--Flow Event Reconstruction in {CMS} and Performance for Jets,
  Taus, and {\MET}'',} CMS Physics Analysis Summary CMS-PAS-PFT-09-001, 2009.

\bibitem{CMS-PAS-PFT-10-001}
\href {http://cdsweb.cern.ch/record/1247373}{{CMS Collaboration},
  ``Commissioning of the particle-flow event reconstruction with the first
  {LHC} collisions recorded in the {CMS} detector'',} CMS Physics Analysis
  Summary CMS-PAS-PFT-10-001, 2010.

\bibitem{madgraph5}
J.~Alwall\hrefCMSnoop {}{ {et~al.}, ``{MadGraph 5: going beyond}'',} \textit{
  JHEP} \textbf{ 06} (2011) 128,
  \href{http://dx.doi.org/10.1007/JHEP06(2011)128}{\doi{10.1007/JHEP06(2011)128}},
\href{http://www.arXiv.org/abs/1106.0522}{\texttt{arXiv:1106.0522}}.

\bibitem{powheg}
\hrefCMSnoop {}{S.~Frixione, P.~Nason, and C.~Oleari, ``{Matching NLO QCD
  computations with parton shower simulations: the POWHEG method}'',} \textit{
  JHEP} \textbf{ 11} (2007) 070,
  \href{http://dx.doi.org/10.1088/1126-6708/2007/11/070}{\doi{10.1088/1126-6708/2007/11/070}},
\href{http://www.arXiv.org/abs/0709.2092}{\texttt{arXiv:0709.2092}}.

\bibitem{Nason:2004rx}
\hrefCMSnoop {}{P.~Nason, ``{A New method for combining NLO QCD with shower
  Monte Carlo algorithms}'',} \textit{ JHEP} \textbf{ 11} (2004) 040,
  \href{http://dx.doi.org/10.1088/1126-6708/2004/11/040}{\doi{10.1088/1126-6708/2004/11/040}},
\href{http://www.arXiv.org/abs/hep-ph/0409146}{\texttt{arXiv:hep-ph/0409146}}.

\bibitem{Alioli:2010xd}
\hrefCMSnoop {}{S.~Alioli, P.~Nason, C.~Oleari, and E.~Re, ``{A general
  framework for implementing NLO calculations in shower Monte Carlo programs:
  the POWHEG BOX}'',} \textit{ JHEP} \textbf{ 06} (2010) 043,
  \href{http://dx.doi.org/10.1007/JHEP06(2010)043}{\doi{10.1007/JHEP06(2010)043}},
\href{http://www.arXiv.org/abs/1002.2581}{\texttt{arXiv:1002.2581}}.

\bibitem{Frixione:2007nw}
\hrefCMSnoop {}{S.~Frixione, P.~Nason, and G.~Ridolfi, ``{A positive-weight
  next-to-leading-order Monte Carlo for heavy flavour hadroproduction}'',}
  \textit{ JHEP} \textbf{ 09} (2007) 126,
  \href{http://dx.doi.org/10.1088/1126-6708/2007/09/126}{\doi{10.1088/1126-6708/2007/09/126}},
\href{http://www.arXiv.org/abs/0707.3088}{\texttt{arXiv:0707.3088}}.

\bibitem{pythia}
\hrefCMSnoop {}{T.~Sj{\"o}strand, S.~Mrenna, and P.~Z. Skands, ``{PYTHIA} 6.4
  physics and manual'',} \textit{ JHEP} \textbf{ 05} (2006) 026,
  \href{http://dx.doi.org/10.1088/1126-6708/2006/05/026}{\doi{10.1088/1126-6708/2006/05/026}},
\href{http://www.arXiv.org/abs/hep-ph/0603175}{\texttt{arXiv:hep-ph/0603175}}.

\bibitem{Pumplin:2002vw}
J.~Pumplin\hrefCMSnoop {}{ {et~al.}, ``{New generation of parton distributions
  with uncertainties from global QCD analysis}'',} \textit{ JHEP} \textbf{ 07}
  (2002) 012,
  \href{http://dx.doi.org/10.1088/1126-6708/2002/07/012}{\doi{10.1088/1126-6708/2002/07/012}},
\href{http://www.arXiv.org/abs/hep-ph/0201195}{\texttt{arXiv:hep-ph/0201195}}.

\bibitem{ct10}
H.-L. Lai\hrefCMSnoop {}{ {et~al.}, ``{New parton distributions for collider
  physics}'',} \textit{ Phys. Rev. D} \textbf{ 82} (2010) 074024,
  \href{http://dx.doi.org/10.1103/PhysRevD.82.074024}{\doi{10.1103/PhysRevD.82.074024}},
\href{http://www.arXiv.org/abs/1007.2241}{\texttt{arXiv:1007.2241}}.

\bibitem{geant}
\hrefCMSnoop {}{{GEANT4} Collaboration, ``{GEANT4---a simulation toolkit}'',}
  \textit{ Nucl. Instrum. Meth. A} \textbf{ 506} (2003) 250,
\href{http://dx.doi.org/10.1016/S0168-9002(03)01368-8}{\doi{10.1016/S0168-9002(03)01368-8}}.

\bibitem{xs-1}
\hrefCMSnoop {}{R.~Gavin, Y.~Li, F.~Petriello, and S.~Quackenbush, ``{FEWZ 2.0:
  A code for hadronic Z production at next-to-next-to-leading order}'',}
  \textit{ Comput. Phys. Commun.} \textbf{ 182} (2011) 2388,
  \href{http://dx.doi.org/10.1016/j.cpc.2011.06.008}{\doi{10.1016/j.cpc.2011.06.008}},
\href{http://www.arXiv.org/abs/1011.3540}{\texttt{arXiv:1011.3540}}.

\bibitem{Gavin:2012sy}
\hrefCMSnoop {}{R.~Gavin, Y.~Li, F.~Petriello, and S.~Quackenbush, ``{W Physics
  at the LHC with FEWZ 2.1}'',} \textit{ Comput. Phys. Commun.} \textbf{ 184}
  (2013) 208,
  \href{http://dx.doi.org/10.1016/j.cpc.2012.09.005}{\doi{10.1016/j.cpc.2012.09.005}},
\href{http://www.arXiv.org/abs/1201.5896}{\texttt{arXiv:1201.5896}}.

\bibitem{xs-2}
\hrefCMSnoop {}{J.~M. Campbell, R.~K. Ellis, and C.~Williams, ``{Vector boson
  pair production at the LHC}'',} \textit{ JHEP} \textbf{ 07} (2011) 018,
  \href{http://dx.doi.org/10.1007/JHEP07(2011)018}{\doi{10.1007/JHEP07(2011)018}},
\href{http://www.arXiv.org/abs/1105.0020}{\texttt{arXiv:1105.0020}}.

\bibitem{xs-3}
\hrefCMSnoop {}{N.~Kidonakis, ``{Next-to-next-to-leading-order collinear and
  soft gluon corrections for t-channel single top quark production}'',}
  \textit{ Phys. Rev. D} \textbf{ 83} (2011) 091503,
  \href{http://dx.doi.org/10.1103/PhysRevD.83.091503}{\doi{10.1103/PhysRevD.83.091503}},
\href{http://www.arXiv.org/abs/1103.2792}{\texttt{arXiv:1103.2792}}.

\bibitem{Czakon:2011xx}
\hrefCMSnoop {}{M.~Czakon and A.~Mitov, ``{Top++: A Program for the Calculation
  of the Top-Pair Cross-Section at Hadron Colliders}'',} \textit{ Comput. Phys.
  Commun.} \textbf{ 185} (2014) 2930,
  \href{http://dx.doi.org/10.1016/j.cpc.2014.06.021}{\doi{10.1016/j.cpc.2014.06.021}},
\href{http://www.arXiv.org/abs/1112.5675}{\texttt{arXiv:1112.5675}}.

\bibitem{Bern:2011pa}
Z.~Bern\hrefCMSnoop {}{ {et~al.}, ``Driving missing data at next-to-leading
  order'',} \textit{ Phys. Rev. D} \textbf{ 84} (2011) 114002,
  \href{http://dx.doi.org/10.1103/PhysRevD.84.114002}{\doi{10.1103/PhysRevD.84.114002}},
\href{http://www.arXiv.org/abs/1106.1423}{\texttt{arXiv:1106.1423}}.

\bibitem{sat-llk}
J.~K. Lindsey, ``{Parametric Statistical Inference}''.
\newblock {Oxford University Press}, 1996.
\newblock {ISBN 0-19-852359-9}.

\bibitem{read}
\hrefCMSnoop {}{A.~L. Read, ``Presentation of search results: the {$CL_s$}
  technique'',} \textit{ J. Phys. G} \textbf{ 28} (2002) 2693,
\href{http://dx.doi.org/10.1088/0954-3899/28/10/313}{\doi{10.1088/0954-3899/28/10/313}}.

\bibitem{junk}
\hrefCMSnoop {}{T.~Junk, ``{Confidence level computation for combining searches
  with small statistics}'',} \textit{ Nucl. Instrum. Meth. A} \textbf{ 434}
  (1999) 435,
  \href{http://dx.doi.org/10.1016/S0168-9002(99)00498-2}{\doi{10.1016/S0168-9002(99)00498-2}},
\href{http://www.arXiv.org/abs/hep-ex/9902006}{\texttt{arXiv:hep-ex/9902006}}.

\bibitem{higgs-comb}
\hrefCMSnoop {}{{ATLAS and CMS Collaborations, LHC Higgs Combination Group},
  ``{Procedure for the LHC Higgs boson search combination in Summer 2011}'',}
  (2011). \href{http://cds.cern.ch/record/1379837}{Technical Report
  ATL-PHYS-PUB 2011-11, CMS NOTE 2011/005}.

\bibitem{Beenakker:1996ch}
\hrefCMSnoop {}{W.~Beenakker, R.~H{\"o}pker, M.~Spira, and P.~M. Zerwas,
  ``Squark and gluino production at hadron colliders'',} \textit{ Nucl. Phys.
  B} \textbf{ 492} (1997) 51,
  \href{http://dx.doi.org/10.1016/S0550-3213(97)80027-2}{\doi{10.1016/S0550-3213(97)80027-2}},
\href{http://www.arXiv.org/abs/hep-ph/9610490}{\texttt{arXiv:hep-ph/9610490}}.

\bibitem{PhysRevD.80.095004}
\hrefCMSnoop {}{A.~Kulesza and L.~Motyka, ``{Soft gluon resummation for the
  production of gluino-gluino and squark-antisquark pairs at the LHC}'',}
  \textit{ Phys. Rev. D} \textbf{ 80} (2009) 095004,
  \href{http://dx.doi.org/10.1103/PhysRevD.80.095004}{\doi{10.1103/PhysRevD.80.095004}},
\href{http://www.arXiv.org/abs/0905.4749}{\texttt{arXiv:0905.4749}}.

\bibitem{PhysRevLett.102.111802}
\hrefCMSnoop {}{A.~Kulesza and L.~Motyka, ``Threshold Resummation for
  Squark-Antisquark and Gluino-Pair Production at the {LHC}'',} \textit{ Phys.
  Rev. Lett.} \textbf{ 102} (2009) 111802,
  \href{http://dx.doi.org/10.1103/PhysRevLett.102.111802}{\doi{10.1103/PhysRevLett.102.111802}},
\href{http://www.arXiv.org/abs/0807.2405}{\texttt{arXiv:0807.2405}}.

\bibitem{1126-6708-2009-12-041}
W.~Beenakker\hrefCMSnoop {}{ {et~al.}, ``Soft-gluon resummation for squark and
  gluino hadroproduction'',} \textit{ JHEP} \textbf{ 12} (2009) 041,
  \href{http://dx.doi.org/10.1088/1126-6708/2009/12/041}{\doi{10.1088/1126-6708/2009/12/041}},
\href{http://www.arXiv.org/abs/0909.4418}{\texttt{arXiv:0909.4418}}.

\bibitem{doi:10.1142/S0217751X11053560}
W.~Beenakker\hrefCMSnoop {}{ {et~al.}, ``Squark and gluino hadroproduction'',}
  \textit{ Int. J. Mod. Phys. A} \textbf{ 26} (2011) 2637,
  \href{http://dx.doi.org/10.1142/S0217751X11053560}{\doi{10.1142/S0217751X11053560}},
\href{http://www.arXiv.org/abs/1105.1110}{\texttt{arXiv:1105.1110}}.

\bibitem{susy-nlo-nll}
M.~Kr{\"a}mer\hrefCMSnoop {}{ {et~al.}, ``{Supersymmetry production cross
  sections in pp collisions at $\sqrt{s} = 7\TeV$}'',} (2012).
\href{http://www.arXiv.org/abs/1206.2892}{\texttt{arXiv:1206.2892}}.

\bibitem{fastsim}
\hrefCMSnoop {}{{CMS Collaboration}, ``The fast simulation of the {CMS}
  detector at {LHC}'',} \textit{ J. Phys. Conf. Ser.} \textbf{ 331} (2011)
  032049,
  \href{http://dx.doi.org/10.1088/1742-6596/331/3/032049}{\doi{10.1088/1742-6596/331/3/032049}}.

\bibitem{lep1}
\hrefCMSnoop {}{{LEP2 SUSY working group (ALEPH, DELPHI, L3 and OPAL
  experiments)}, ``{Combined LEP Chargino Results, up to 208 GeV for large
  m$_0$}'',} (2001). \href{http://lepsusy.web.cern.ch/lepsusy}{Note
  LEPSUSYWG/01-03.1}.

\bibitem{lep2}
\hrefCMSnoop {}{{LEP2 SUSY working group (ALEPH, DELPHI, L3 and OPAL
  experiments)}, ``{Combined LEP Chargino Results, up to 208 GeV for low
  DM}'',} (2004). \href{http://lepsusy.web.cern.ch/lepsusy}{Note
  LEPSUSYWG/02-04.1}.

\bibitem{Perelstein:2008zt}
\hrefCMSnoop {}{M.~Perelstein and A.~Weiler, ``{Polarized Tops from Stop Decays
  at the LHC}'',} \textit{ JHEP} \textbf{ 03} (2009) 141,
  \href{http://dx.doi.org/10.1088/1126-6708/2009/03/141}{\doi{10.1088/1126-6708/2009/03/141}},
\href{http://www.arXiv.org/abs/0811.1024}{\texttt{arXiv:0811.1024}}.

\end{thebibliography}\endgroup
\cleardoublepage \appendix\section{The CMS Collaboration \label{app:collab}}\begin{sloppypar}\hyphenpenalty=5000\widowpenalty=500\clubpenalty=5000\textbf{Yerevan Physics Institute,  Yerevan,  Armenia}\\*[0pt]
V.~Khachatryan, A.M.~Sirunyan, A.~Tumasyan
\vskip\cmsinstskip
\textbf{Institut f\"{u}r Hochenergiephysik der OeAW,  Wien,  Austria}\\*[0pt]
W.~Adam, E.~Asilar, T.~Bergauer, J.~Brandstetter, E.~Brondolin, M.~Dragicevic, J.~Er\"{o}, M.~Flechl, M.~Friedl, R.~Fr\"{u}hwirth\cmsAuthorMark{1}, V.M.~Ghete, C.~Hartl, N.~H\"{o}rmann, J.~Hrubec, M.~Jeitler\cmsAuthorMark{1}, A.~K\"{o}nig, M.~Krammer\cmsAuthorMark{1}, I.~Kr\"{a}tschmer, D.~Liko, T.~Matsushita, I.~Mikulec, D.~Rabady, N.~Rad, B.~Rahbaran, H.~Rohringer, J.~Schieck\cmsAuthorMark{1}, J.~Strauss, W.~Treberer-Treberspurg, W.~Waltenberger, C.-E.~Wulz\cmsAuthorMark{1}
% ...
\vskip\cmsinstskip
% ...
1:~~Also at Vienna University of Technology, Vienna, Austria\\
% ...

\end{sloppypar}
\end{document}